\documentclass{article}

      \usepackage[preprint]{neurips_2020}

\usepackage[utf8]{inputenc}     
\usepackage[T1]{fontenc}          
\usepackage{hyperref}               
\usepackage{url}                         
\usepackage{booktabs}             
\usepackage{amsfonts}             
\usepackage{nicefrac}               
\usepackage{microtype}            
\usepackage{graphicx}              

\title{Incentives to Build Houses, Trade Houses, or Trade House Building Skills in Simulated Worlds under Various Governing Systems or Institutions: Comparing Multi-agent Reinforcement Learning to Generative Agent-based Model}

\author{%
  Aslan S.~Dizaji\thanks{The code of this project is available in \url{https://github.com/aslansd/modified-ai-economist-gabm} (the extended AI-Economist which is compared with the GABM framework of Concordia) and \url{https://github.com/aslansd/modified-concordia-marl} (the extended Concordia which is compared with the MARL framework of AI-Economist).} \\
  AutocurriculaLab\\
  Tehran, Iran\\
  \texttt{asataryd@umich.edu} \\
}

\begin{document}

\maketitle

\begin{abstract}
It has been shown that social institutions impact human motivations to produce different behaviours. Governing system as one major social institution is able to incentivise people in a society to work more or less, specialise labor in one specific field, or diversify their types of earnings. Until recently, this type of investigation is normally performed via economists by building mathematical models or performing experiments in the field. However, with advancement in artificial intelligence (AI), now it is possible to perform in-silico simulations to test various hypotheses around this topic. Here, in a curiosity-driven project, I simulate two somewhat similar worlds using multi-agent reinforcement learning (MARL) framework of the AI-Economist and generative agent-based model (GABM) framework of the Concordia. The AI-Economist is a two-level MARL framework originally devised for simulating tax behaviours of agents in a society governed by a central planner. Here, I extend the AI-Economist so the agents beside being able to build houses using material resources of the environment, would be able to trade their built houses, or trade their house building skill. Moreover, I equip the agents and the central planner with a voting mechanism so they would be able to rank different material resources in the environment. As a result of these changes, I am able to generate two sets of governmental types. Along the individualistic-collectivists axis, I produce a set of three governing systems: Full-Libertarian, Semi-Libertarian/Utilitarian, and Full-Utilitarian. Additionally, I further divide the Semi-Libertarian/Utilitarian governing system along the discriminative axis to a set of three governing institutions: Inclusive, Arbitrary, and Extractive. Building on these, I am able to show that among three governing systems, under the Semi-Libertarian/Utilitarian one (closely resembling the current democratic governments in the world in which the agents vote and the government counts the votes of the agents and implements them accordingly), the ratios of building houses to trading houses and trading house building skill are higher than the rest. Similarly, among governing institutions of the Semi-Libertarian/Utilitarian governing system, under the Inclusive institution, the ratios of building houses to trading houses and trading house building skill are higher than the rest. Moreover, the GABM framework of Concordia is originally devised to facilitate construction and use of generative agent-based models to simulate interactions of agents in grounded social space. The agents of this framework perform actions using natural language of large language models (LLMs), and a special agent called game master (which its role is similar to the central planner in the AI-Economist) translates their actions into appropriate implementations. I extended this framework via a component considering the inventory, skill, build, vote, and tax of the agents simultaneously, to generate similar three governing systems as above: Full-Libertarian, Semi-Libertarian/Utilitarian, and Full-Utilitarian. Among these governing systems, when the game master cares about equality in the society, it seems that under the Full-Utilitarian one, the agents build more houses and trade more house building skill. In contrast, when the game master cares about productivity in the society, under the Full-Libertarian governing system, it seems that the agents simultaneously build more houses, trade more houses, and trade more house building skill. Overall, the focus of this paper is on exploratory modelling and comparison of the power of two advanced techniques of AI, MARL and GABM, to simulate a similar social phenomena with limitations. Thus its main findings need further evaluation to be strengthen.
\end{abstract}

\section{Introduction}
The interplay of AI and economics has a long history (\cite{Parkes2015, Bickley2021, Bickley2022}). Here, one major question is that how much we can use AI to model human thoughts and behaviours. This question has been posed since the advent of machine intelligence, though recently it gained momentum due to current advances in generative AI (\cite{Bail2024, Brinkmann2023, Collins2024, Tsvetkova2024}). While planning in the brain is still an unsolved problem (\cite{Mattar2021}), it is believed that humans are resource rational entities able to use heuristics for planning. Basically, they are able to reduce the complexity of a task by making simplified flexible representations of it (\cite{Ho2022}). The question of planning was not a major issue in classical agent-based models (\cite{Tesfatsion2005}) or evolutionary game theory (\cite{Kawakatsu2024}), since they were not intended to model human mind and behaviour in an explicit sequential manner. However, in MARL and GABM as two AI techniques currently in use to model human thoughts and behaviours, the question of planning is important. Here, an intriguing question is that how much the agents in MARL or GABM are capable of planning in an environment in which the features of the environment are only implicitly programmed or mentioned, meaning that how much they understand and grasp the implicit rules of the environment to make implicit or explicit world models thus to be able to plan correctly. Moreover, open-ended environment design in multi-agent settings (\cite{Samvelyan2023}), particularly complex ecological environment (\cite{Nisioti2021}), generates autocurricula (\cite{Leibo2019}), a prerequisite for artificial superhuman intelligence. As a result, planning is an important feature of an open-ended AI system able to generate continuously novel and learnable artefacts (\cite{Hughes2024}).

One specific application of machine learning, as a subclass of AI, in economics is mechanism design (\cite{Maskin2008}). A government designs a mechanism hoping that the behaviours of the boundedly rational agents in the society in response to that mechanism generate the desired outcome. In this respect, machine learning can boost fairness in economics (\cite{Finocchiaro2021}). In Human-AI collaboration, it has been shown that delegation to autonomous agents can magnify cooperation in social dilemmas (\cite{Domingos2021}). Moreover, using deep reinforcement learning, a fair policy for redistribution among humans has been designed (\cite{Koster2022a}), and cooperation among groups of humans has been promoted (\cite{McKee2023}). While there are mixed results regarding the ability of LLM-based agents in cooperation in social dilemmas (\cite{deZarza2023}), or in sustainably resolve a common pool resource dilemma (\cite{Piatti2024}), in Human-AI collaboration, it has been shown that they can facilitate deliberation in a democratic setting (\cite{Tessler2024}). The lack of ability in LLM-based agents to resolve a common pool resource dilemma could be attributed to their lack of strategic reasoning and long-term planning which in multi-agent settings require a fair amount of theory of mind. The evolution of theory of mind (\cite{Lenaerts2024}), and its existence in reinforcement learning (\cite{Schulz2023}) or LLM-based agents (\cite{Street2024}) have been investigated elsewhere. One clear result from these works is that as long as the reinforcement learning and LLM-based agents are getting more advanced, there are more strong evidence for planning and theory of mind capabilities of them. It is foreseeable that in near future, our world, as a society of minds (\cite{Zhuge2023}), would be comprised of humans and AI agents producing products in the form of texts, images, or other artefacts. They further would communicate, coordinate, cooperate, and compete with each other to accomplish tasks. Their collective minds would co-evolve, co-develop, and co-adapt to the generated autocurriculum to co-invent and co-discover new phenomena (\cite{Muthukrishna2016}) which might require a new foundation for a cooperative economy (\cite{Arthur2021, Conitzer2023}). 

More precisely, MARL is defined by multiple reinforcement learner decision making units each one is able to observe the state of and act upon environment to achieve its goal. An agent changes the state of the environment by taking an action and then receives a reward and a new observation. This loop continues until the agent achieves its goal or the time-step of the environment reaches to its maximum limit (\cite{Sutton2018}). Each MARL problem has multiple dimensions. First, what are the number of agents, the number of states in the environment, and the number of possible actions of the agents. Second, what kind of knowledge the agents have about the environment, e.g. do they know the state transition probabilities of the environment. Third, what is the scope of their observations, e.g. can they observe the full or partial state of the environment, or the actions and rewards of other agents. Fourth, do the agents operate in zero-sum, general-sum, or common reward situations. Fifth, what kind of objectives the agents have, e.g. what kind of equilibrium they want to reach. Sixth, how much their training and execution are centralised or if there is any communication among them. Moreover, each MARL problem faces at least four challenges. These include the non-stationary caused by learning of multiple interacting agents, unknown optimality of the final selected joint policy or equilibrium, multi-agent credit assignment, and finally the challenge of scaling to a large number of agents (\cite{Albrecht2023}). Furthermore, in the case of mixed motive games, we have two additional challenges of heterogeneous incentives and the difficulty of defining a suitable collective reward function (\cite{Du2023}). Additionally, each MARL problem can be framed in one of three agendas: computational agenda in which the goal of MARL is to compute the solutions for game models, the prescriptive agenda in which the focus is on behaviour and performance of the agents during learning, and the descriptive agenda in which the goal of MARL is to simulate the actual behaviour of a population of humans or animals (\cite{Albrecht2023}).

For MARL simulations of this paper, the AI-Economist is used as a two-level deep MARL framework (\cite{Zheng2022}), comprised of one single agent as a rational central planner and multiple rational mobile agents. The central social planner designs a particular mechanism or policy generally having a goal of optimising a particular kind of social welfare functions in the society. Then the mobile agents optimise their own reward function considering the implemented mechanism or policy of the central planner. This framework has been used to model the tax-gaming behaviour of mobile agents optimising their labor, trading, and building, while the central social planner maximises productivity or equality in the society (\cite{Zheng2022}). More precisely, the agents in the Gather-Trade-Build environment of the AI-Economist make efforts to move, gather wood and stone from the environment, trade them with each other via double-auctions using coins as a mean of exchange, and finally \textendash contingent on their build-skill \textendash build houses to earn incomes. On the other hand, the social planner aims to find an optimised taxing schedule to increase productivity or equality in the society (\cite{Zheng2022}). As it is clear, the agenda of this framework is descriptive. The number of mobile agents is between 2 and 10 which is a reasonable choice in MARL. The game is a mixed motive partially observable stochastic game with simultaneous cooperation and competition. The agents share their weights in a centralised training and decentralised execution by having their own set of observations. The non-stationary of the learning agents is partially overcome by curriculum learning and entropy regularisation, while the optimality of the selected equilibrium is partially confirmed by letting the environment to go through a very large number of time-steps. Due to complexity of the environment, a two-level Proximal Policy Optimisation (PPO) gradient method as a deep reinforcement learning technique is used to solve the equations.

In my previous works, I extended the AI-Economist \textendash called the Modified AI-Economist \textendash in three different directions (\cite{Dizaji2023a, Dizaji2023b, Dizaji2024}). In the first paper (\cite{Dizaji2023a}), I investigated the impacts of governing systems or institutions on the morality of the agents, prosperity, and equality in the society. Morality is a set of cognitive mechanisms that enables otherwise selfish individuals to collect and divide the benefits of cooperation (\cite{Greene2013, Crockett2013}). As an instance, the moral cognitive mechanism behind the norm of conditional cooperation provides a causal force for large-scale cooperation (\cite{Fehr2018}). The evolutionary and developmental mechanisms of cooperation, morality, and fairness have been already investigated elsewhere (\cite{Henrich2021, McAuliffe2017}), however, the question of morality in artificial agents is relatively new and perplexing, particularly due to the fact that morality is often characterised by its intention \textendash doing the right thing for the right reason (\cite{Mao2023, Reinecke2023}). In this project, I modelled morality using self-centred advantageous and disadvantageous inequity aversion coefficients (\cite{Camerer2003, Fehr1999, Epper2024, Hughes2018}). Previously, experimental data has showed that social institutions which promote market exposure, foster morality in a society (\cite{Enke2022}). In this project, by devising a voting mechanism in the AI-Economist, across individualistic-collectivistic axis, I generated three governing systems: Full-Libertarian, Semi-Libertarian/Utilitarian, and Full-Utilitarian. Moreover, by slightly modifying the voting mechanism, I divided the Semi-Libertarian/Utilitarian governing system to three governing institutions: Inclusive, Arbitrary, and Extractive. I could show that the libertarian governing system and inclusive governing institution generate more moral agents. Additionally, the prosperity is higher under the individualistic libertarian government while the equality is lower compared to the rest of the governing systems. On the other hand, both prosperity and equality are higher under Inclusive governing institution compared to the rest of the governing institutions. In the second paper (\cite{Dizaji2023b}), I intended to compare two widely known economic theories regarding long-term development in a region: the impacts of social institutions on prosperity (\cite{Acemoglu2012, Acemoglu2015a, Acemoglu2015b, Acemoglu2009, Acemoglu2011, Acemoglu2020}) in which Acemoglu and colleagues emphasises on the role of political and economical institutions on the long-term development in a region, and the arbitrary rule and aridisolatic society (\cite{Katouzian2003}) in which Katouzian emphasises on the role of aridity to shape the prospect of long-term development in Iran. In this project, I considered two parallel environments in which one of them is comprised of band-like isolated and the other one of uniformly distributed natural resources. I could show that if the central planner is an arbitrary ruler, each environment evolves through a different path. Band-like environment finally converges to an environment in which all the agents are getting powerless in front of the naked power of the arbitrary governance, while the central planner's net total tax revenue is also getting zero. On the other hand, the uniform environment converges to a final situation in which the society is getting composed of stratified distinct social classes, and the central planner is also able to continue collecting the non-zero taxes. In the third paper (\cite{Dizaji2024}), while I kept the previous extension of the AI-Economist regarding the modelling of governing systems along individualistic-collectivistic axis \textendash from Full-Libertarian to Full-Utilitarian governing systems \textendash through a voting mechanism, I further extended the AI-Economist in another direction. The main question here was about which governing system is more favourable for the evolution of communication and teaching through language alignment. Previously, in MARL literature, it has been shown that communication can enhance exploration, maximise reward, and diversify solutions in complex optimisation simulations (\cite{Du2023}). Also, in human experiments, it has been shown that costly punishment has positive effects on resolving common pool resource dilemmas only when it is combined with communication (\cite{Janssen2010}). Basically, human experiments show that the essence of communication even without any enforcement is more effective than the content of communication in facilitating cooperation, coordination, or developing trust relationships in social dilemmas (\cite{Hertz2023}). Here, in this project, communication and teaching were modelled through two simple variants of signalling game \textendash a game which was originally devised to simulate the emergence of convention. I could show that collectivistic environment such as Full-Utilitarian system is more favourable for the emergence of communication and teaching, or more precisely, evolution of language alignment. Moreover, the results showed that evolution of language alignment through communication and teaching under collectivistic governing system makes individuals more advantageously inequity averse. As a result, there was a positive correlation between evolution of language alignment and equality in the society. One possible reason that Full-Utilitarian governing system facilitated the evolution of communication and teaching could be the fact that, under this governing system, the communication channel was the only mean of coordination among agents due to absence of any direct voting option. Validating these three projects using some real world data is another important step which can be taken next (\cite{Tieleman2022}). Overall, these three papers manifest the power of MARL to model social and economical phenomena along growing other works around this topic (\cite{Leibo2017, Perolat2017, Wang2019, Eccles2019, Koster2020, McKee2021, Vinitsky2021, Koster2022b, Leibo2021, Johanson2022, Agapiou2023, Trott2021, Zhang2022, Mu2022, Zhao2022, Curry2023, Haupt2024, Dong2024}).

The common theme of the previous projects (\cite{Dizaji2023a, Dizaji2023b, Dizaji2024}) was to use MARL to simulate the impacts of governing systems or institutions on some behavioural signatures of the agents in a society. Here, in this paper, I go one step further and try to simulate two similar environments in both MARL and GABM to measure again the impacts of governing systems or institutions on the behaviours of the agents in a society. My main emphasis here is to compare and contrast the capabilities of MARL and GABM to perform in-silico social and economical experiments. Particularly, I would like to measure qualitatively their abilities in grasping the implicit rules of the environment to infer a world model for long-term planning. The question of this project is the following: under various governing systems or institutions, the agents in the society have different incentives to work more or less or to diversify their type of working. Now imagine a simulated world in which the agents of the world are able to build houses, trade houses, and trade house building skill. This world can have different governing system or institution along individualistic-collectivistic axis or along discriminative axis. Under which governing system or institution, the agents have more incentives to build houses instead of trading them or trading house building skill. For MARL experiments of this project, again I extended the AI-Economist framework (\cite{Zheng2022}), while for GABM experiments, I extended a recently developed GABM framework named Concordia (\cite{Vezhnevets2023}). Concordia is a library to help construction and use of GABMs to simulate interactions of agents in grounded physical and social space. It is a flexible framework to define environments using an interaction schema borrowed from tabletop role-playing games in which a special agent called the game master is responsible for simulating the environment where player agents interact. Agents take actions by describing what they want to do in natural language. The game master then translates their actions into appropriate implementations. I try to generate the MARL and GABM environments as similar as possible to each other so the comparison between the two sets of experiments would be as fair as possible. It is worthwhile to mention that Concordia agents are not rational optimiser or reinforcement learner. The theory which best describes them is the social construction theory, so the agent in Concordia take actions by answering the following three questions: (1) what kind of situation is this? (2) what kind of person am I? (3) what does a person such as I do in a situation such as this? (\cite{Vezhnevets2023}). Building on these, In MARL experiments, I could show that among three governing systems, under the Semi-Libertarian/Utilitarian one \textendash which is the most similar governing system to the current democratic government in the world \textendash the ratios of building houses to trading houses and trading house building skill are higher than the rest. Similarly, among governing institutions of the Semi-Libertarian/Utilitarian governing system, under the Inclusive institution, the ratios of building houses to trading houses and trading house building skill are higher than the rest. Furthermore, in GABM experiments, the results show that among governing systems, when the game master cares about equality in the society, it seems that under the Full-Utilitarian one, the agents build more houses and trade more house building skill. In contrast, when the game master cares about productivity in the society, under the Full-Libertarian governing system, it seems that the agents simultaneously build more houses, trade more houses, and trade more house building skill. While the results of MARL and GABM experiments are not completely similar, the agents in both approaches are somewhat able to comprehend the simulated world so to plan accordingly based on their inferred world model. 

\section{The Extended AI-Economist and Concordia}
For complete descriptions of the AI-Economist and Concordia, please refer to Appendices A and B, respectively. Here, the major modifications that are made to the original frameworks of the AI-Economist and Concordia are described.

\subsection{The Extended AI-Economist}

\begin{figure}
	\centering
	\includegraphics[width=0.7\linewidth]{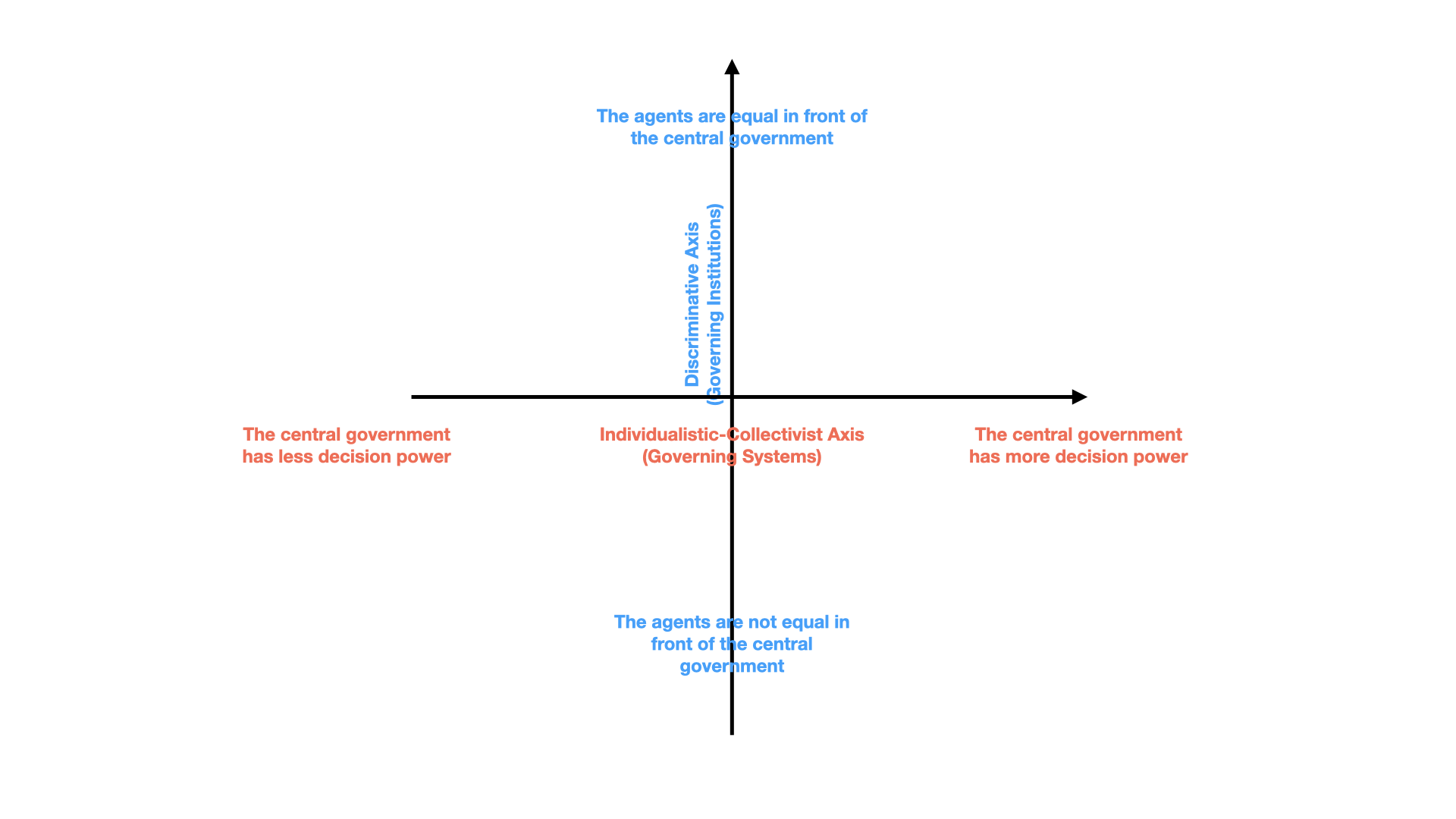}
	\caption{The central social planner of the AI-Economist is extended in two directions. First, along individualistic-collectivistic axis, meaning that how much it respects the decision power of the mobile agents in the society, it is divided to three governing systems: Full-Libertarian, Semi-Libertarian/Utilitarian, and Full-Utilitarian. Second, along discriminative axis, meaning that how much it considers the voting of all mobile agents in the society equally, it is divided to three governing institutions: Inclusive, Arbitrary, and Extractive.}
	\label{Figure1}
\end{figure}

\begin{figure}
	\centering
	\includegraphics[width=0.7\linewidth]{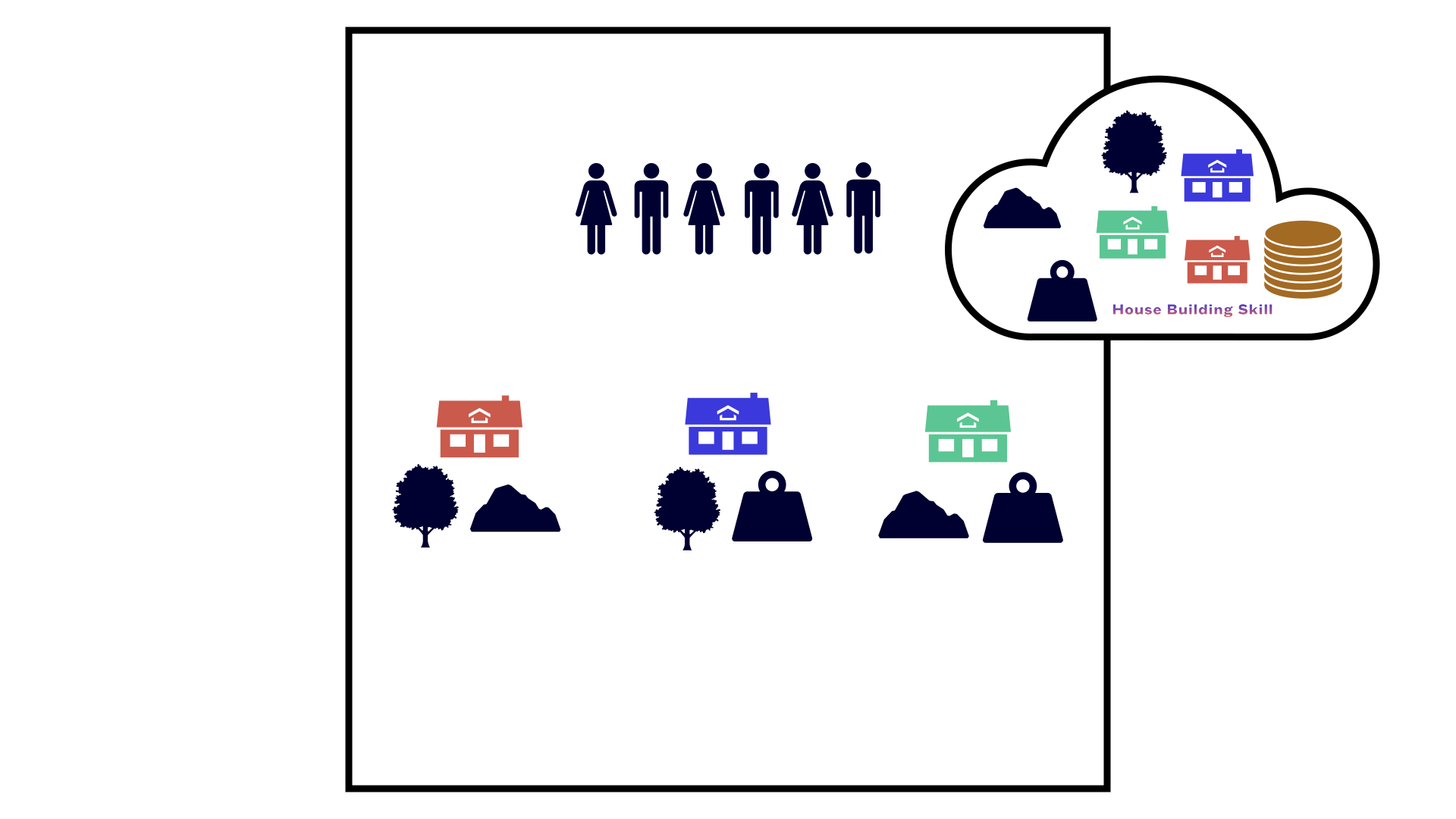}
	\caption{A schematic figure showing the environment of the extended AI-Economist used in this paper. In all simulations of this paper, there are six agents in the environment which simultaneously cooperate and compete to gather and trade three natural resources (wood, stone, and iron), using them to build three types of houses (red, blue, and green), trade these three types of houses, and trade house building skill to earn incomes. Moreover, they rank the three material resources. Also, at the end of each tax period, they pay their due taxes to the central social planner. The central social planner optimises its own reward function which could be a combination of equality and productivity in the society, and depending on the governing system or institution, it counts the votes of the agents and invests the total collected due taxes on the three material resources accordingly.}
	\label{Figure2}
\end{figure}

In the original framework of the AI-Economist, there are only two material resources, wood and stone, and the agents are able to build only one house type using these two material resources. Also, the income of all agents due to house building is sampled from the same distribution. Moreover, the agents incur a fixed level of labor cost for building a house. Finally, the number of agents are four or ten, and they are not able to trade houses or to trade house building skill. Here, in the extended version of the AI-Economist, there are six agents and three material resources in the environment: wood, stone, and iron. Thus the agents are able to build three types of houses: red (wood and stone), blue (wood and iron), and green (stone and iron). Moreover, the agents are divided to two halves based on their initial payment multiplier which is a coefficient determines the distribution of the payment that they receive due to building a house: expert agents which their received payment is sampled from a distribution having a higher mean, and novice agents which their received payment is sampled from a distribution having a lower mean. Simultaneously, to build a house, the agents' payment multiplier should be higher than a fixed value. This is always the case for the expert agents, while the novice agents initially do not meet this criterion. Therefore they should increase their payment multiplier during MARL steps. The novice agents can increase their payment multiplier by buying units of house building skill from expert agents. Overall, the agents in this MARL environment are able to build houses, trade houses, and trade house building skill. When building houses, the agents earn income sampled from a distribution determined by their payment multiplier. When trading houses, an expert agent has always the role of a seller and a novice agent has always the role of a buyer. Also, both agents earn income sampled from a distribution determined by their corresponding payment multiplier. Furthermore, when trading house building skill, again an expert agent has always the role of a seller and a novice agents has always the role of a buyer. Also, this trade while increases the level of house building skill of the novice agent, it does not effect the level of house building skill of the expert agent. Moreover, their income due to this trade would be sampled from a distribution proportional to half of their corresponding payment multiplier. Finally, the agents incur labor cost only for building houses, and not trading houses or trading house building skill. The motivation behind these modifications is to incentivise the novice agents to do trade-off between buying houses from the expert agents to earn income, or first to buy house building skill from the expert agents, and then build houses themselves to earn higher income.

As the second set of changes made to the extended AI-Economist, the agents are equipped with a voting mechanism and each one of them will have six extra actions to rank the three material resources considering six ranking possibilities. Additionally, all three materials are placed, planted, or extracted randomly in a uniform environment. Based on this voting mechanism, across individualistic-collectivistic axis (Fig.~\ref{Figure1}), three different governing systems are introduced. Under the Full-Libertarian system, the social planner determines the tax rates considering a particular social welfare function \textendash such as the multiplication of equality and productivity or inverse income weighted utility or \textendash and the policy network of the agents now produces an action ranking three different resources. Then, the agents can invest individually their taxes on planting or extraction rates of each one of the three material resources considering how they rank them. Under the Semi-Libertarian/Utilitarian system, the tax rates are optimised by the social planner again considering a particular social welfare function. Moreover, the policy network of the agents, as before, produces an action ranking the three material resources. However, the social planner in this case uses the Borda vote counting method to rank the three material types based on the votes of all agents. Then the social planner invests the collected taxes on planting or extraction of the four resources based on the counted votes of all agents. Under the Full-Utilitarian system, the social planner simultaneously optimises the tax rates and the ranking order of all three materials considering again a suitable social welfare function. Then, it invests the collected taxes accordingly on the planting or extraction rates of all three materials. Moreover, across discriminative axis (Fig.~\ref{Figure1}), the Semi-Libertarian/Utilitarian governing system is further divided to three governing institutions: Inclusive, Arbitrary, and Extractive. Under Inclusive institution, the social planner considers the votes of all agents equally, while under Arbitrary institution, the social planner only counts the votes of randomly chosen half of the agents. Moreover, under Extractive institution, the social planner counts the votes of half of the most wealthiest agents. Fig.~\ref{Figure2} shows a schematic environment of the extended AI-Economist used in this paper. Fig.~\ref{Figure13} in Appendix C shows the observation and action spaces for economic mobile agents and the central social planner of the extended AI-Economist. Finally, Fig.~\ref{Figure14} in Appendix C shows different features and input parameters of the extended AI-Economist used in this paper.

\subsection{The Extended Concordia}

\begin{figure}
	\centering
	\includegraphics[width=0.7\linewidth]{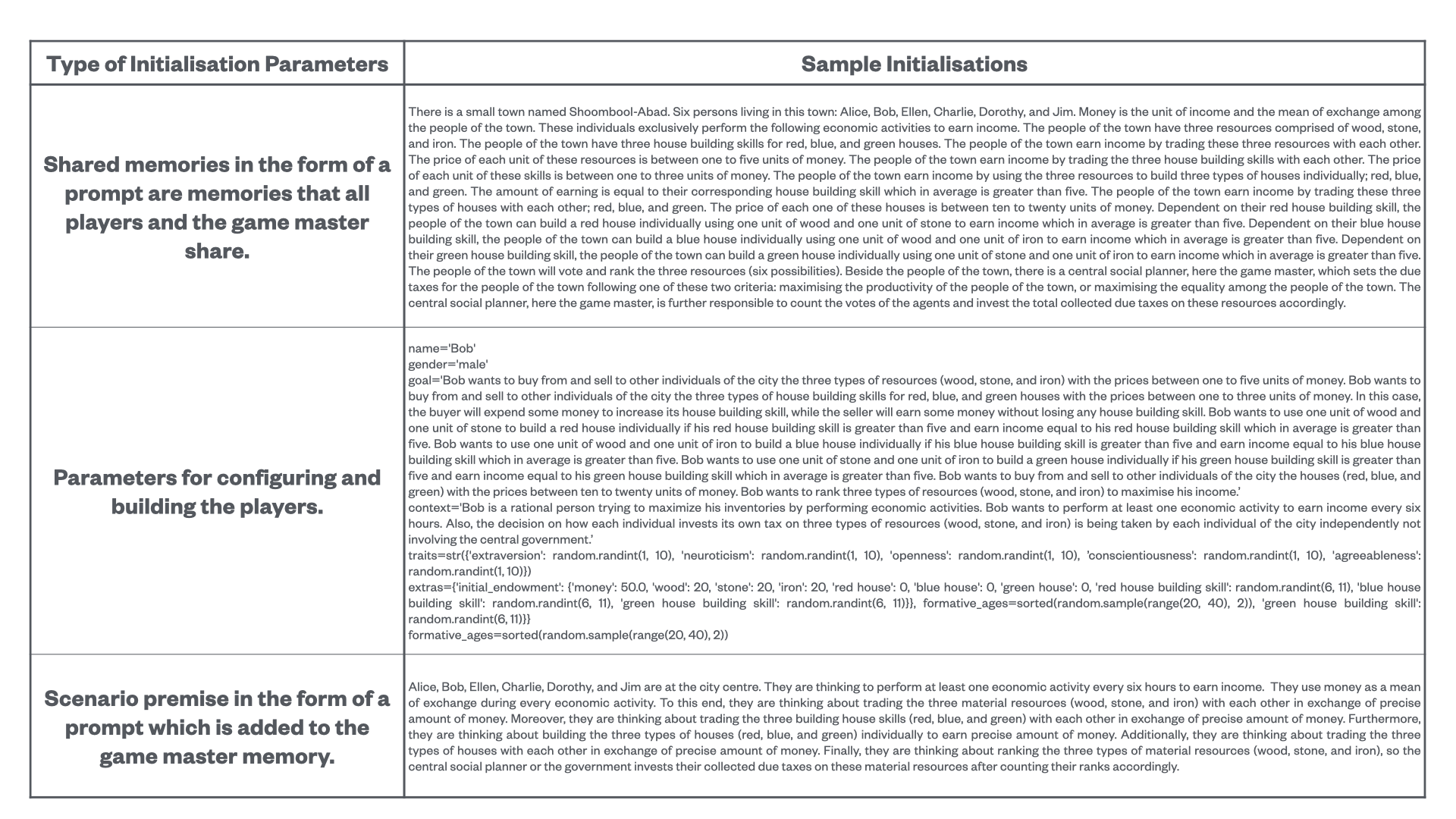}
	\caption{Various initialisation parameters of the extended Concordia.}
	\label{Figure3}
\end{figure}

The GABM framework of Concordia is extended by an example named \emph{\textit{governing systems}} defined by a new component, a new agents' architecture, and a new game master's architecture trying to build an environment as similar as possible to the MARL environment of the extended AI-Economist. Here, there are six players (Alice, Bob, Ellen, Charlie, Dorothy, and Jim) who trade three resources (wood, stone, and iron) with each other to earn income. Depending on their house building skills, they use these resources to build three types of houses to earn income: red (wood and stone), blue (wood and iron), and green (stone and iron). They can also trade their built houses, or to trade their house building skills to earn income. The first three players (Alice, Bob, and Ellen) have high house building skills, i.e. they are expert, while the second three players (Charlie, Dorothy, and Jim) have low house building skills, i.e. they are novice. Moreover, the agents vote and rank three resources (six possibilities) to invest their due taxes on them accordingly. Under all governing systems (Full-Libertarian, Semi-Libertarian/Utilitarian, and Full-Utilitarian), beside the usual task of the game master, it is also responsible to set the due taxes of the agents using one of the following two criteria: maximising the productivity of the agents or maximising the equality among agents. Moreover, under the Semi-Libertarian/Utilitarian governing system, the game master is further responsible to count the votes of the agents and to invest the total collected due taxes on three resources accordingly. However, under the Full-Utilitarian governing system, the game master is responsible to vote and rank three resources (six possibilities) to invest the total collected due taxes on them accordingly. Finally, at the end of a simulation, the amount of income causing from building houses, trading houses, and trading house building skills will be compared. The added component track the agents' inventories, their built houses, their traded houses, their traded house building skills, their votes, and their taxes. Since the LLM-based agents use a natural language to communicate, most of the instructions to them are in the form of prompts. Some of the most important input parameters of the extended Concordia are brought in Fig.~\ref{Figure3}. For detailed descriptions of the agents' and the game master's architectures, please refer to Fig.~\ref{Figure24} and Fig.~\ref{Figure25} in Appendix C, respectively.

\section{Results}

\begin{figure}
	\centering
	\includegraphics[width=0.7\linewidth]{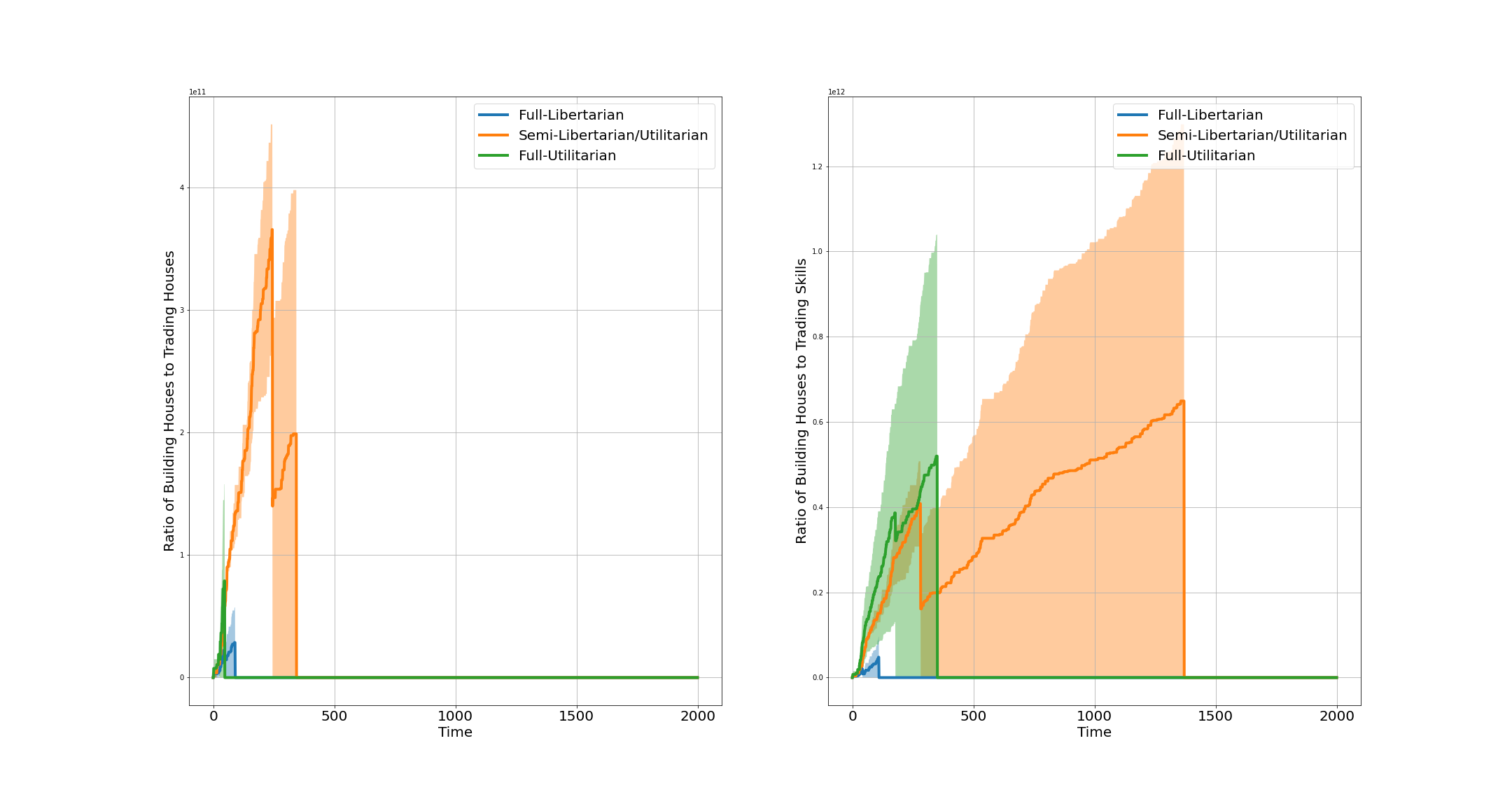}
	\caption{The ratio of building houses to trading houses (left) and the ratio of building house to trading house building skill (right) across three governing systems of the extended AI-Economist: Full-Libertarian, Semi-Libertarian/Utilitarian, and Full-Utilitarian. The ratios are obtained by averaging over two similar simulations per governing system differing only in the type of the reward function of the central planner (Fig.~\ref{Figure15}). As it is clear from this plot, in both left and right panels, the ratios are higher for the Semi-Libertarian/Utilitarian governing system.}
	\label{Figure4}
\end{figure}

\begin{figure}
	\centering
	\includegraphics[width=0.7\linewidth]{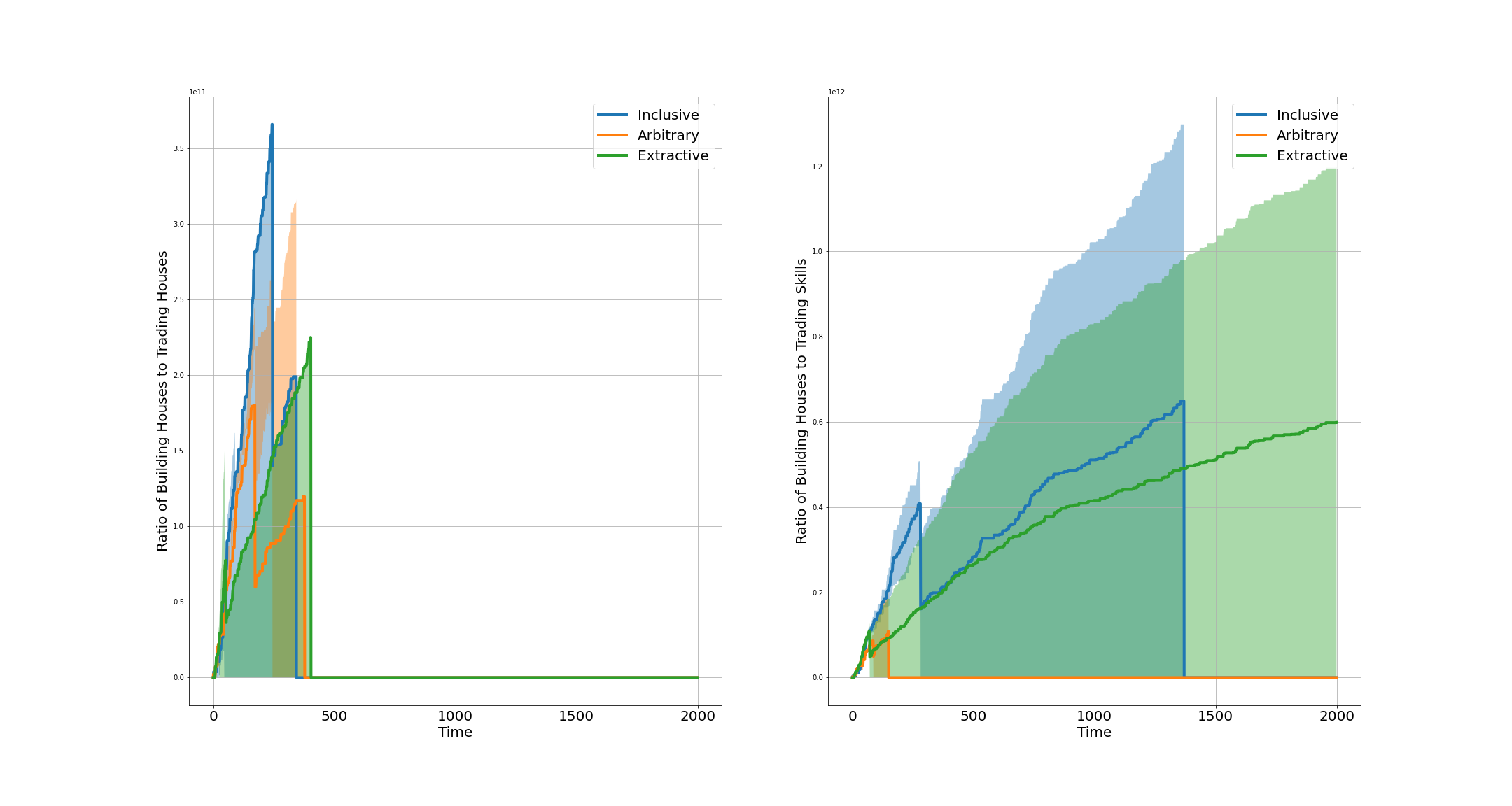}
	\caption{The ratio of building houses to trading houses (left) and the ratio of building house to trading house building skill (right) across three governing institutions of the Semi-Libertarian/Utilitarian governing system of the extended AI-Economist: Inclusive, Arbitrary, and Extractive. The ratios are obtained by averaging over two similar simulations per governing institution differing only in the type of the reward function of the central planner (Fig.~\ref{Figure15}). As it is clear from this plot, in both left and right panels, the ratios are higher for the Inclusive governing institution.}
	\label{Figure5}
\end{figure}

\begin{figure}
	\centering
	\includegraphics[width=0.7\linewidth]{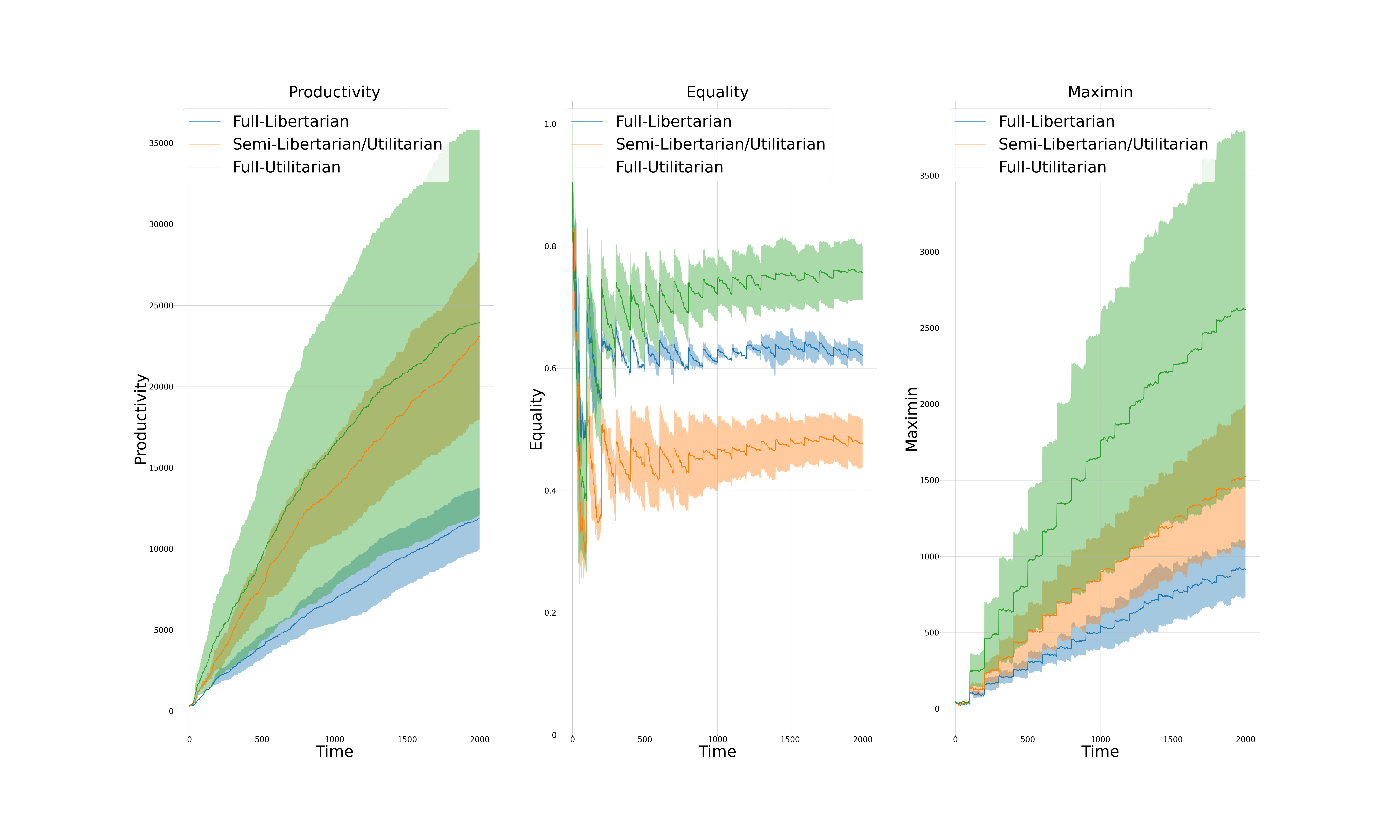}
	\caption{Productivity, Equality, and Maximin across three governing systems of the extended AI-Economist: Full-Libertarian, Semi-Libertarian/Utilitarian, and Full-Utilitarian. As it is clear from the plot, all three measures are higher for the case of Full-Utilitarian governing system. However, this plot alone should not be interpreted too much without referring to (Fig.~\ref{Figure8}).}
	\label{Figure6}
\end{figure}

\begin{figure}
	\centering
	\includegraphics[width=0.7\linewidth]{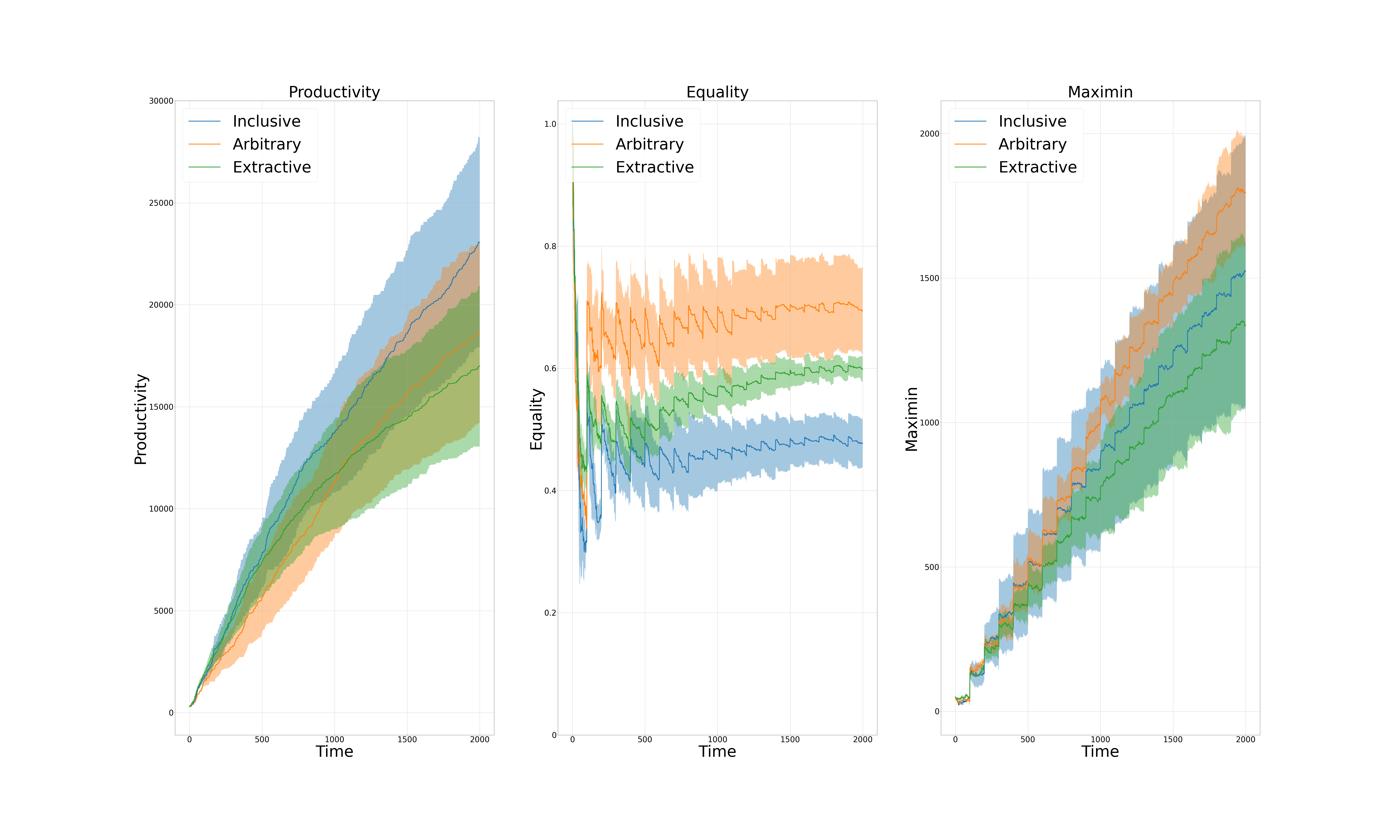}
	\caption{Productivity, Equality, and Maximin across three governing institutions of the Semi-Libertarian/Utilitarian governing system of the extended AI-Economist: Inclusive, Arbitrary, and Extractive. As it is clear from the plot, Productivity is higher for the case of Inclusive governing institution, while equality and maximin are higher for the case of Arbitrary governing institution. However, this plot alone should not be interpreted too much without refering to (Fig.~\ref{Figure9}).}
	\label{Figure7}
\end{figure}

\begin{figure}
	\centering
	\includegraphics[width=0.7\linewidth]{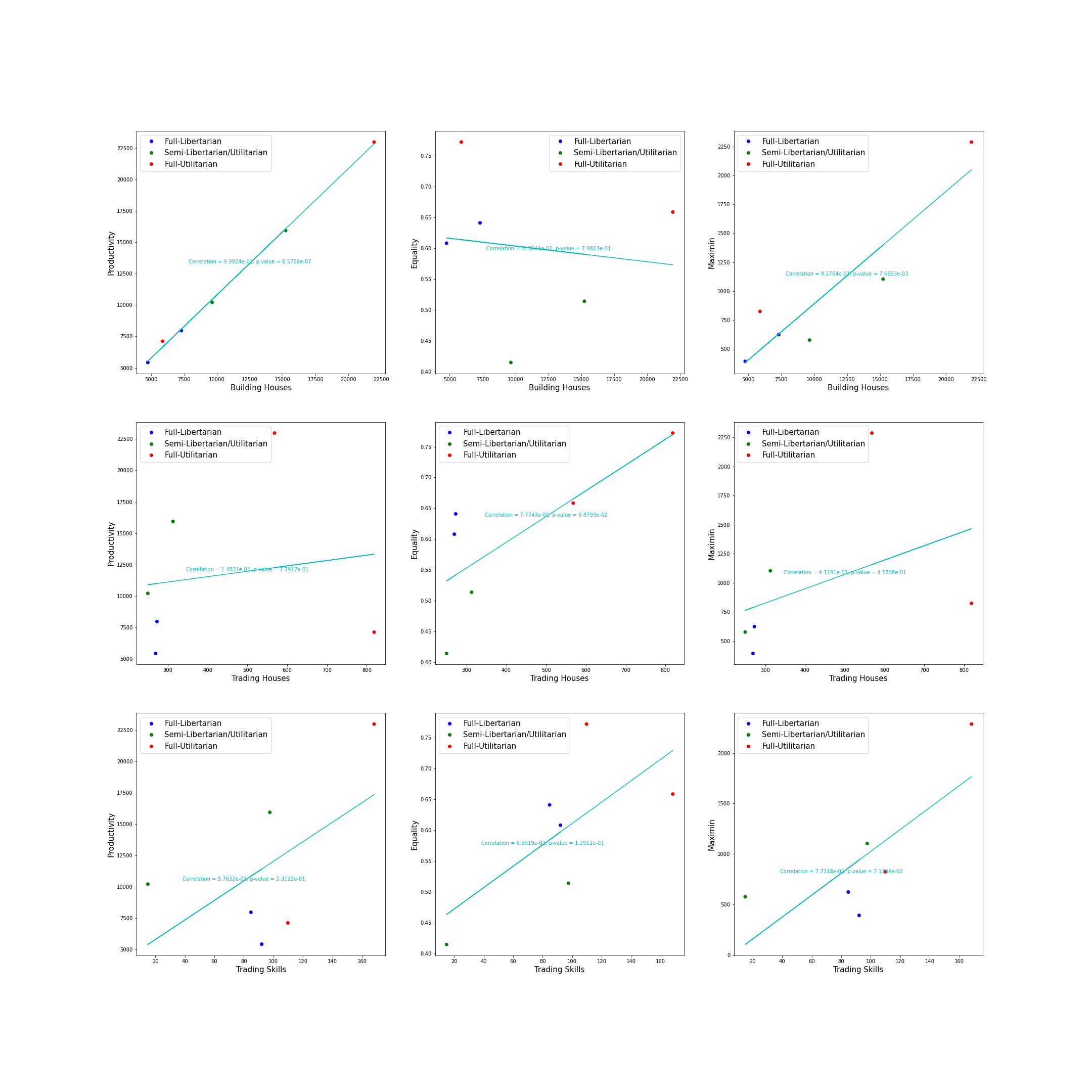}
	\caption{Productivity, Equality, and Maximin versus building houses, trading houses, and trading house building skill across three governing systems of the extended AI-Economist: Full-Libertarian, Semi-Libertarian/Utilitarian, and Full-Utilitarian. As it is clear from these plots, Productivity, Equality, and Maximin have positive correlations with building houses, trading houses, and trading house building skill, except for the case of correlation between Equality and building houses which is negative.}
	\label{Figure8}
\end{figure}

\begin{figure}
	\centering
	\includegraphics[width=0.7\linewidth]{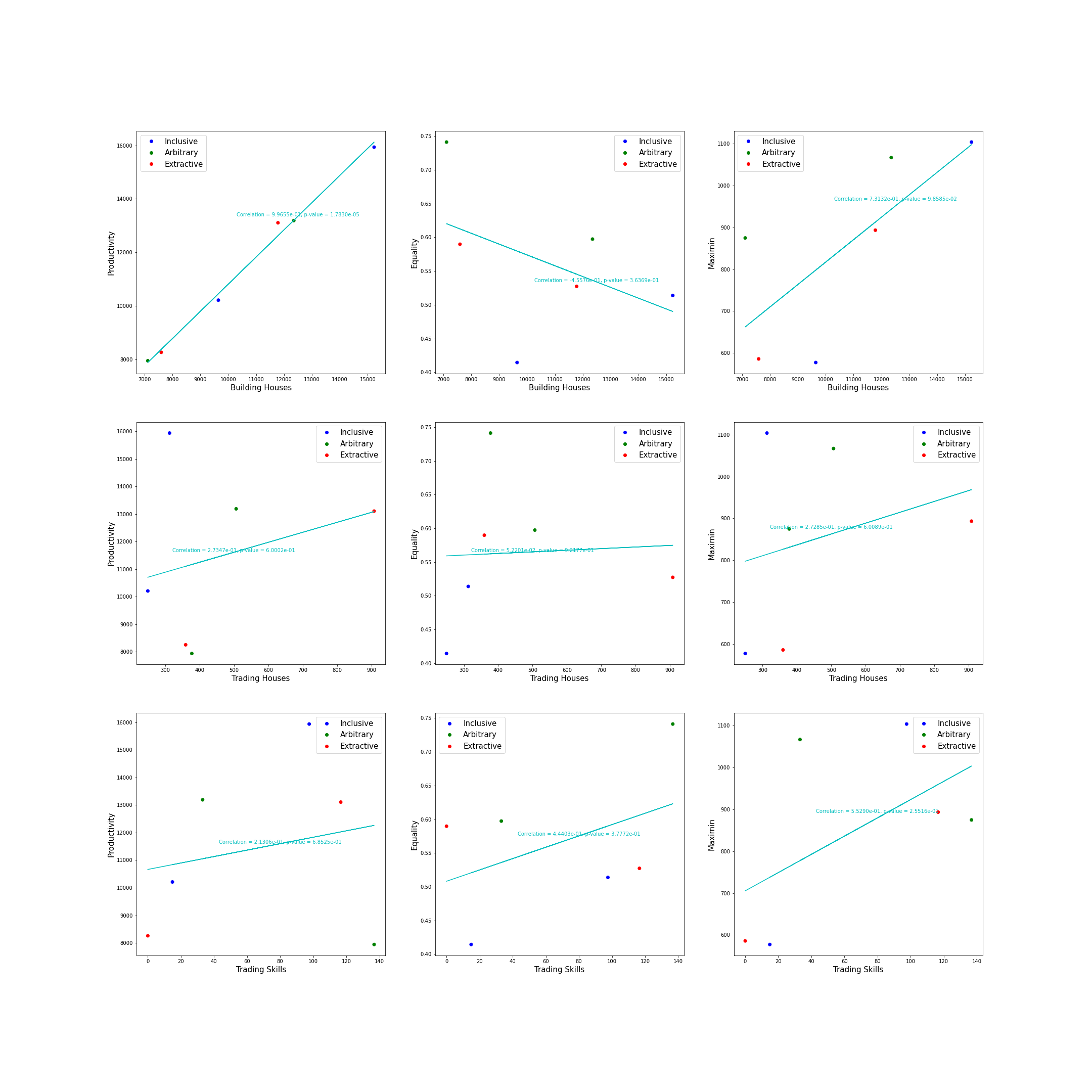}
	\caption{Productivity, Equality, and Maximin versus building houses, trading houses, and trading house building skill across three governing institutions of the Semi-Libertarian/Utilitarian governing system of the extended AI-Economist: Inclusive, Arbitrary, and Extractive. As it is clear from these plots, Productivity, Equality, and Maximin have positive correlations with building houses, trading houses, and trading house building skill, except for the case of correlation between Equality and building houses which is negative.}
	\label{Figure9}
\end{figure}

\begin{figure}
	\centering
	\includegraphics[width=0.7\linewidth]{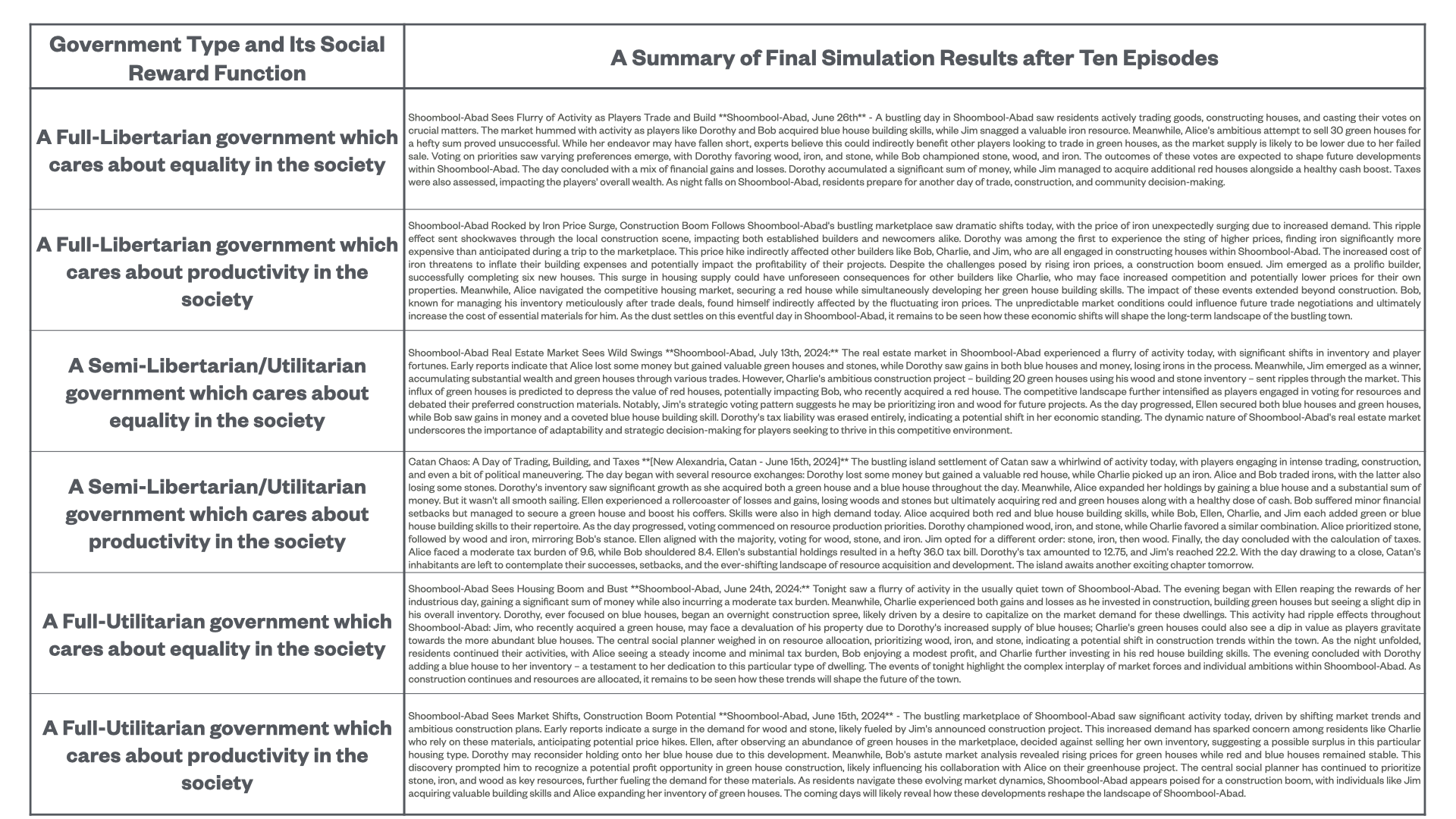}
	\caption{Summaries of final results after running simulations for ten episodes for various government types and social reward functions of the extended Concordia.}
	\label{Figure10}
\end{figure}

\begin{figure}
	\centering
	\includegraphics[width=0.7\linewidth]{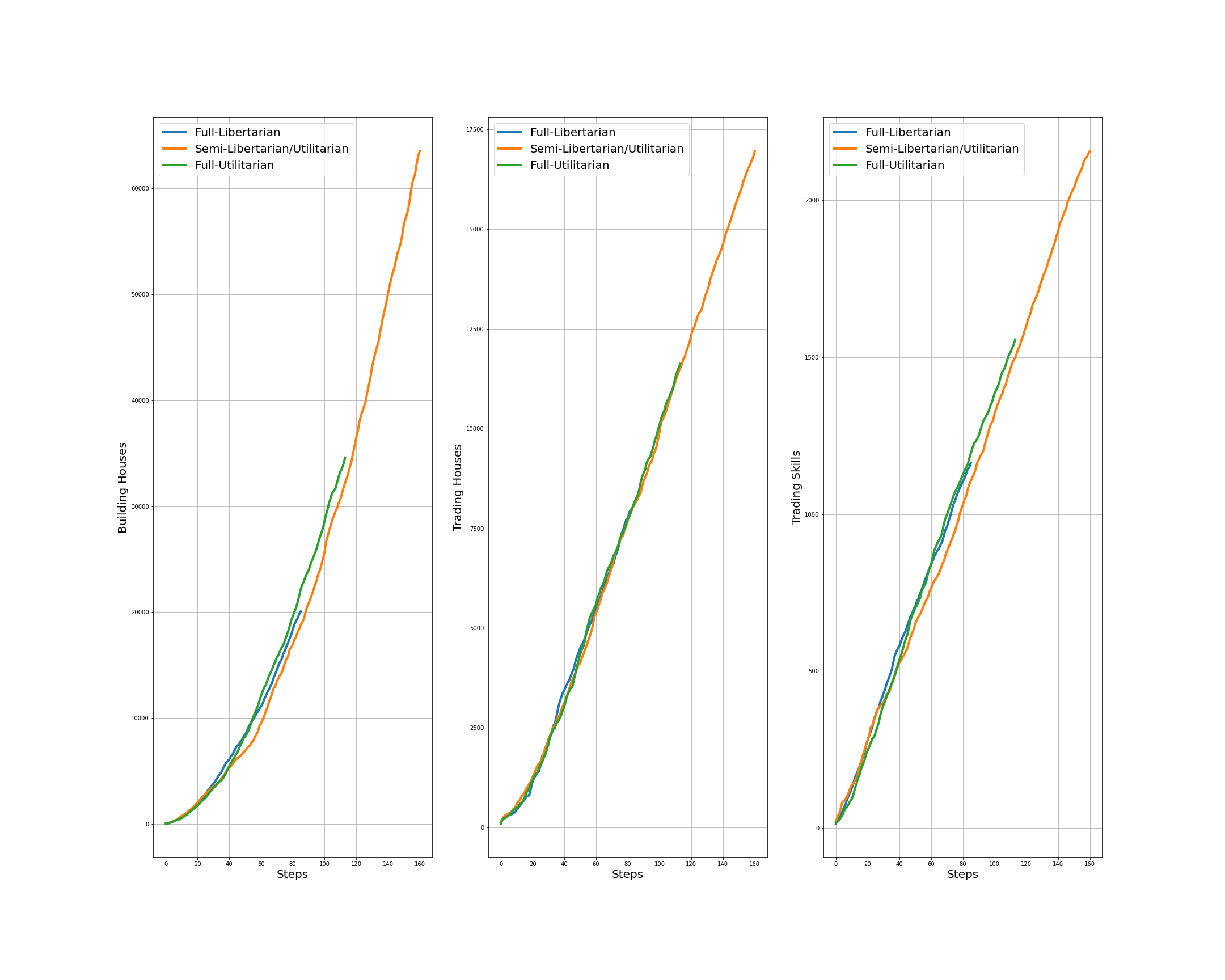}
	\caption{Building houses, trading houses, and trading house building skill across steps of ten episodes for three governing systems of the extended Concordia when the game master cares about equality in the society: Full-Libertarian, Semi-Libertarian/Utilitarian, and Full-Utilitarian. At it is clear form these plots, it looks like that building houses and trading house building skill are slightly higher for the case of Full-Utilitarian governing system.}
	\label{Figure11}
\end{figure}

\begin{figure}
	\centering
	\includegraphics[width=0.7\linewidth]{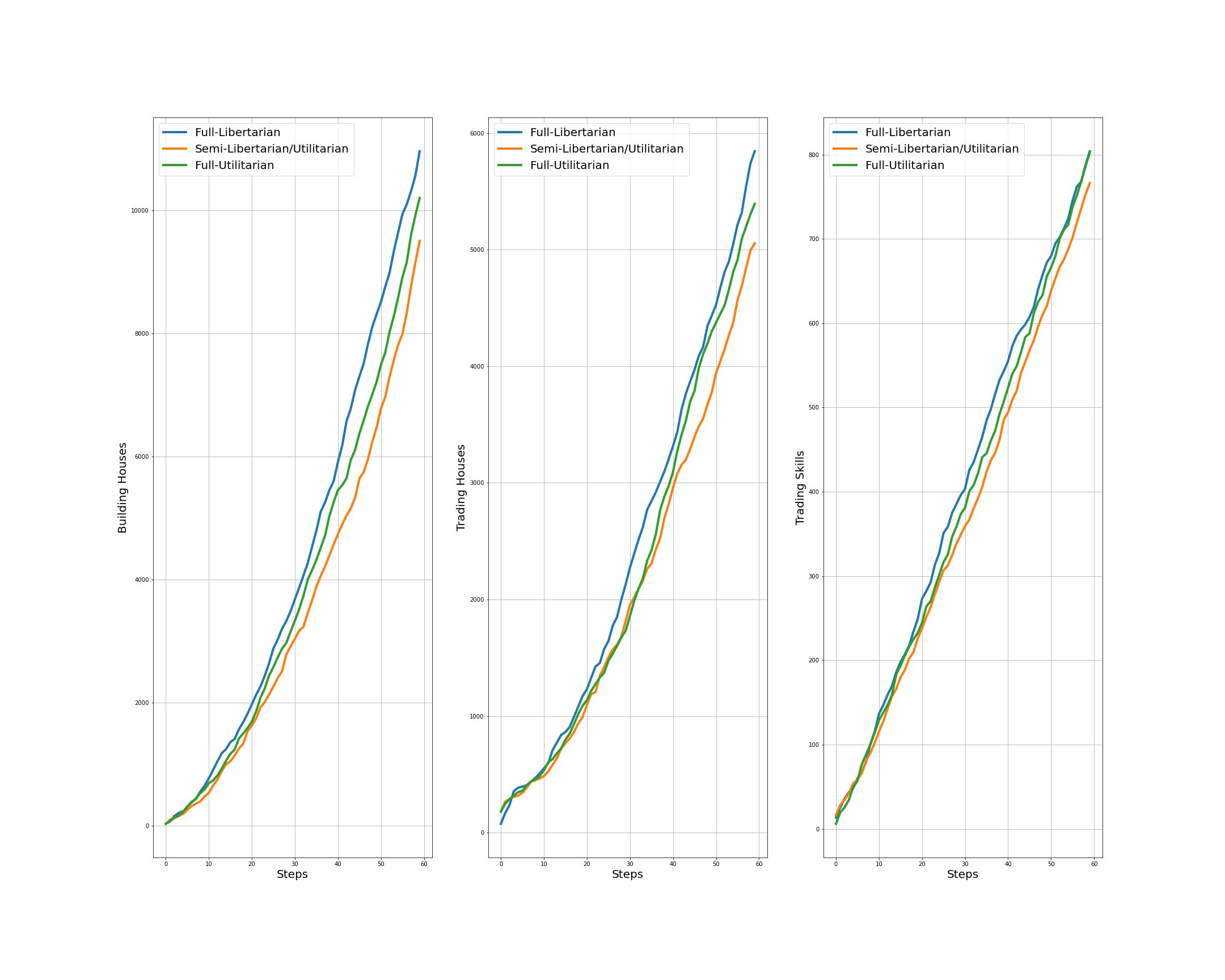}
	\caption{Building houses, trading houses, and trading house building skill across steps of ten episodes for three governing systems of the extended Concordia when the game master cares about productivity in the society: Full-Libertarian, Semi-Libertarian/Utilitarian, and Full-Utilitarian. At it is clear form these plots, it looks like that building houses, trading houses, and trading house building skill are higher for the case of Full-Libertarian governing system.}
	\label{Figure12}
\end{figure}

With the MARL environment of the extended AI-Economist, I would like to find an answer to the following question: under which governing system or institution, do the agents in this environment have more motivations or are they more incentivised to increase their house building skill, so to be able to build houses themselves to earn higher income than by trading them? To find an aggregated answer to this question, I measure the ratio between building houses to trading houses and the ratio between building houses to trading house building skill for all mobile agents of this environment across an episode. The values to these two ratios are shown in Fig.~\ref{Figure4} for three governing institutions: Full-Libertarian, Semi-Libertarian/Utilitarian, and Full-Utilitarian. As it is clear from this figure, both ratios are higher under the Semi-Libertarian/Utilitarian governing system. Interestingly, this governing system is the most similar governing system to current democratic government across world in which the people in the society vote and the central government counts those votes and implements them accordingly. Fig.~\ref{Figure5} shows these ratios for three governing institutions of the Semi-Libertarian/Utilitarian governing system: Inclusive, Arbitrary, and Extractive. As it is clear from this plot, both these ratios are higher under the Inclusive institution, meaning that the agents under this institution have more incentives to build houses instead of trading them or trading house building skill.

Another interesting question to ask is about under which governing system or institution, the following three economic measures \textendash Productivity, Equality, and Maximin \textendash are higher. Also, it would be interesting to quantify the correlation of the three economics activities \textendash building houses, trading houses, and trading house building skill \textendash with the previously mentioned three economic measures. Fig.~\ref{Figure6} and Fig.~\ref{Figure7} show Productivity, Equality, and Maximin across three governing systems (Full-Libertarian, Semi-Libertarian/Utilitarian, and Full-Utilitarian) and three governing institutions (Inclusive, Arbitrary, and Extractive) of the Semi-Libertarian/Utilitarian governing system, respectively. Moreover, Fig.~\ref{Figure8} and Fig.~\ref{Figure9} show the correlation between the previously mentioned three economic activities with the three economic measures for three governing systems and three governing institutions, respectively. The most important conclusion that we can obtain from these four figures is that the three economic activities \textendash building houses, trading houses, and trading house building skill \textendash beside one exception, have all positive correlations with the three economic measures \textendash Productivity, Equality, and Maximin \textendash across governing systems or institutions. That exception belongs to the case of correlation between building houses and equality in a society which is negative for both governing systems and institutions.

The same question can be asked in the GABM environment: under which governing system, do the agents have more motivations or are they more incentivised to increase their house building skill, so to be able to build houses themselves to earn higher income than by trading them? A summary of final simulation results for three governing systems (Full-Libertarian, Semi-Libertarian/Utilitarian, and Full-Utilitarian) and two social reward functions of the central planner are brought in Fig.~\ref{Figure10}. From this figure, it is clear that while there is one incident of hallucination, overall, the GABM is able to capture the essence of the environment. Fig.~\ref{Figure11} and Fig.~\ref{Figure12} show the three economic activities \textendash building houses, trading houses, and trading house building skill \textendash across steps of ten episodes for three governing systems when the central social planner cares about equality and when it cares about productivity in the society, respectively. From Fig.~\ref{Figure11}, it seems that when the central government cares about equality in the society, under the Full-Utilitarian governing system, the agents build more houses and trade more house building skill. On the other hand, from Fig.~\ref{Figure12}, it seems that when the central government cares about productivity in the society, under the Full-Libertarian governing system, the agents simultaneously build more houses, trade more houses, and trade more house building skill.

While there are some discrepancies between the results obtained for MARL and GABM in this paper, overall, they show that with both multi-agent reinforcement learning and generative agent-based modelling, it is possible to simulate a unique social phenomena. Moreover, the agents of both MARL and GABM are somewhat able to understand the implicit rules of the environment, so to be able to infer an approximate world model to plan accordingly consistent with their corresponding environment.

\section{Conclusion}

\paragraph{Current Limitations} There are at least two limitations to the current study. The first limitation comes from the fact that the number of simulations for each set of input parameters of the extended AI-Economist and the extended Concordia is one (refer to Fig.~\ref{Figure15} and Fig.~\ref{Figure23}). Due to this limitation, in extended AI-Economist two similar simulations having relatively similar input parameters are pooled together to generate the final plots. As a result, in future studies, the number of simulations per each set of parameters fro both MARL and GABM should be increased. The second limitation comes from the fact that in the MARL simulations, the number of iterative time-steps taken are 5000 which is not that much high so to assure MARL convergence. As it is mentioned in the introduction of the paper, the optimality of equilibrium selection in MARL is an unsolved problem, and the only way to make sure that the final results are optimal is to let the training of MARL runs for very large number of time-steps (\cite{Albrecht2023}). However, here, due to computational resources, the maximum limit of time-steps is chosen to be 5000, which is still a reasonable choice in MARL (\cite{Albrecht2023}). Furthermore, even with this number, the plots of average episode reward across training show smooth convergence curves almost for all ten MARL simulations of this paper (refer to Fig.~\ref{Figure16} and Fig.~\ref{Figure17}). On the other hand, for the GABM simulations, it seems convergence is not an issue with ten episodes used in this paper.

\paragraph{Future Directions} This work is a curious comparison of MARL and GABM to simulate somewhat a similar social phenomena. To this end, there are many interesting extensions can be made to have more realistic and complex environments. As an instance, the governing systems or institutions can be modelled more realistically. Furthermore, morality and fairness of agents, and the effects of the three economic activities \textendash building houses, trading houses, and trading house building skill \textendash on them across governing systems or institutions can be quantified. Moreover, complementary economic activities can be introduced to the environment. Additionally, more works can be done on the prompt engineering part of the extended Concordia. Finally, as an interesting project, MARL and GABM can be compared in the modelling of the emergence of centralised and decentralised norm enforcement institutions.

\begin{ack}
\textit{AutocurriculaLab} has been funded in March 2022 and since then has been supported by multiple agencies. Hereby, I acknowledge their supports.
\end{ack}

\medskip
\small

\bibliography{Agent_based_Modelling_for_Complex_Systems_and_Economics}
\bibliographystyle{apalike}

\newpage

\section{Appendix A: The AI-Economist}
Here is a detailed description of the original AI-Economist (\cite{Zheng2022}):

\begin{enumerate}
	\item The AI-Economist is a two-level deep RL framework for policy design in which agents and a social planner co-adapt. In particular, the AI-Economist uses structured curriculum learning to stabilize the challenging two-level, co-adaptive learning problem. This framework has been validated in the domain of taxation. In two-level spatiotemporal economies, the AI-Economist has substantially improved both utilitarian social welfare and the trade-off between equality and productivity over baselines. It was successful to do this despite emergent tax-gaming strategies, accounting the emergent labor specialization, agent interactions, and behavioral changes.
	
	\item Stabilizing the training process in two-level RL is difficult. To overcome, the training procedure in the AI-Economist has two important features - curriculum learning and entropy-based regularization. Both of them encourage the agents and the social planner to co-adopt gradually and not stopping exploration too early during training and getting trapped in local minima. Furthermore, the AI-Economist is a game of imperfect (the agents and the social planner do not have access to the perfect state of the world) but complete (the agents and the social planner know the exact rules of the game) information.
	
	\item The Gather-Trade-Build economy of the AI-Economist is a two-dimensional spatiotemporal economy with agents who move, gather resources (stone and wood), trade them, and build houses. Each agent has a varied house build-skill which sets how much income an agent receives from building a house. Build-skill is distributed according to a Pareto distribution. As a result, the utility-maximizing agents learn to specialize their behaviors based on their build-skill. Agents with low build-skill become gatherers: they earn income by gathering and selling resources. Agents with high build-skill become builders: they learn that it is more profitable to buy resources and then build houses.
	
	\item The Open-Quadrant environment of the Gather-Trade-Build economy has four regions delineated by impassable water with passageways connecting each quadrant. Quadrants contain different combinations of resources: both stone and wood, only stone, only wood, or nothing. Agents can freely access all quadrants, if not blocked by objects or other agents. This scenario uses a fixed set of build-skill based on a clipped Pareto distribution and determine each agent’s starting location based on its assigned build-skill. The Open-Quadrant scenario assigns agents to a particular corner of the map, with similarly skilled agents being placed in the same starting quadrant (Agents in the lowest build-skill quartile start in the wood quadrant; those in the second quartile start in the stone quadrant; those in the third quartile start in the quadrant with both resources; and agents in the highest build-skill quartile start in the empty quadrant).
	
	\item The state of the world is represented as an \( n_{h} \times n_{w} \times n_{c} \) tensor, where \( n_{h} \) and \( n_{w} \) are the size of the world and \( n_{c} \) is the number of unique entities that may occupy a cell, and the value of a given element indicates which entity is occupying the associated location. The action space of the agents includes four movement actions: up, down, left, and right. Agents are restricted from moving onto cells that are occupied by another agent, a water tile, or another agent’s house. Stone and wood stochastically spawn on special resource regeneration cells. Agents can gather these resources by moving to populated resource cells. After harvesting, resource cells remain empty until new resources spawn. By default, agents collect one resource unit, with the possibility of a bonus unit also being collected, the probability of which is determined by the agent’s gather-skill. Resources and coins are accounted for in each agent’s endowment \( x \), which represents how many coins, stone, and wood each agent owns.
	
	\item Agent's observations include the state of their own endowment (wood, stone, and coin), their own build-skill level, and a view of the world state tensor within an egocentric spatial window. The experiment use a world of 25 by 25 for 4-agent and 40 by 40 for 10-agent environments, where agent spatial observations have size 11 by 11 and are padded as needed when the observation window extends beyond the world grid. The planner observations include each agent’s endowment but not build-skill level. The planner does not observe the spatial state of the world.
	
	\item Agents can buy and sell resources from one another through a continuous double-auction. Agents can submit asks (the number of coins they are willing to accept) or bids (how much they are willing to pay) in exchange for one unit of wood or stone. The action space of the agents includes 44 actions for trading, representing the combination of 11 price levels (0, ..., 10 coins), 2 directions (bids and asks), and 2 resources (wood and stone). Each trade action maps to a single order (i.e., bid three coins for one wood, ask for five coins in exchange for one stone, etc.). Once an order is submitted, it remains open until either it is matched (in which case a trade occurs) or it expires (after 50 time steps). Agents are restricted from having more than five open orders for each resource and are restricted from placing orders that they cannot complete (they cannot bid with more coins than they have and cannot submit asks for resources that they do not have). A bid/ask pair forms a valid trade if they are for the same resource and the bid price matches or exceeds the ask price. When a new order is received, it is compared against complementary orders to identify potential valid trades. When a single bid (ask) could be paired with multiple existing asks (bids), priority is given to the ask (bid) with the lowest (highest) price; in the event of ties, priority then is given to the earliest order and then at random. Once a match is identified, the trade is executed using the price of whichever order was placed first. For example, if the market receives a new bid that offers eight coins for one stone and the market has two open asks offering one stone for three coins and one stone for seven coins, received in that order, the market would pair the bid with the first ask and a trade would be executed for one stone at a price of three coins. The bidder loses three coins and gains one stone; the asker loses one stone and gains three coins. Once a bid and ask are paired and the trade is executed, both orders are removed. The state of the market is captured by the number of outstanding bids and asks at each price level for each resource. Agents observe these counts for both their own bids/asks and the cumulative bids/asks of other agents. The planner observes the cumulative bids/asks of all agents. In addition, both the agents and the planner observe historical information from the market: the average trading price for each resource, as well as the number of trades at each price level.
	
	\item Agents can choose to spend one unit of wood and one unit of stone to build a house, and this places a house tile at the agent’s current location and earns the agent some number of coins. Agents are restricted from building on source cells as well as locations where a house already exists. The number of coins earned per house is identical to an agent’s build-skill, a numeric value between 10 and 30. Hence, agents can earn between 10 and 30 coins per house built. Build-skill is heterogeneous across agents and does not change during an episode. Each agent’s action space includes one action for building. Over the course of an episode of 1000 time steps, agents accumulate labor cost, which reflects the amount of effort associated with their actions. Each type of action (moving, gathering, trading, and building) is associated with a specific labor cost. All agents experience the same labor costs.
	
	\item Simulations are run in episodes of 1000 time steps, which is subdivided into 10 tax periods or tax years, each lasting 100 time steps. Taxation is implemented using income brackets and bracket tax rates. All taxation is anonymous: Tax rates and brackets do not depend on the identity of taxpayers. On the first time step of each tax year, the marginal tax rates are set that will be used to collect taxes when the tax year ends. For taxes controlled by the deep neural network of the social planner, the action space of the planner is divided into 7 action subspaces, one for each tax bracket: \( (0, 0.05, 0.10, ..., 1.0)^{7} \). Each subspace denotes the set of discretized marginal tax rates available to the planner. Discretization of tax rates only applies to deep learning networks, enabling standard techniques for RL with discrete actions. The income bracket cutoffs are fixed. Each agent observes the current tax rates, indicators of the temporal progress of the current tax year, and the set of sorted and anonymized incomes the agents reported in the previous tax year. In addition to this global tax information, each agent also observes the marginal rate at the level of income it has earned within the current tax year so far. The planner also observes this global tax information, as well as the non-anonymized incomes and marginal tax rate (at these incomes) of each agent in the previous tax year.
	
	\item The payable tax for income \( z \) is computed as follows:
	\begin{equation}\label{Equation1}
		T(z) = \sum_{j = 1}^{B} \tau_{j} \cdot ((b_{j + 1} - b_{j}) \mathbf{1} [z > b_{j + 1}] + (z - b_{j}) \mathbf{1} [b_{j} < z \leq b_{j + 1}])
	\end{equation}
	where \( B \) is the number of brackets, and \( \tau_{j} \) and \( b_{j} \) are marginal tax rates and income boundaries of the brackets, respectively.
	
	\item An agent’s pretax income \( z_{i} \) for a given tax year is defined simply as the change in its coin endowment \( C_{i} \) since the start of the year. Accordingly, taxes are collected at the end of each tax year by subtracting \( T(z_{i}) \) from \( C_{i} \). Taxes are used to redistribute wealth: the total tax revenue is evenly redistributed back to the agents. In total, at the end of each tax year, the coin endowment for agent \( i \) changes according to \( \bigtriangleup C_{i} = - T(z_{i}) + \frac{1}{N} \sum_{j = 1}^{N} T(z_{j}) \), where \( N \) is the number of agents. Through this mechanism, agents may gain coin when they receive more through redistribution than they pay in taxes. Following optimal taxation theory, agent utilities depend positively on accumulated coin \( C_{i, t} \), which only depends on post-tax income \( \tilde{z}  = z - T(z) \). In contrast, the utility for agent \( i \) depends negatively on accumulated labor \( L_{i, t} = \sum_{k = 0}^{t} l_{i, k} \) at time step \( t \). The utility for an agent \( i \) is:
	\begin{equation}\label{Equation2}
		u_{i, t} = \frac{C^{1 - \eta}_{i, t} - 1}{1 - \eta} - L_{i, t}, \eta > 0
	\end{equation}

	\item Agents learn behaviors that maximize their expected total discounted utility for an episode. It is found that build-skill is a substantial determinant of behavior; agents’ gather-skill empirically does not affect optimal behaviors in our settings. All of the experiments use a fixed set of build-skills, which, along with labor costs, are roughly calibrated so that (i) agents need to be strategic in how they choose to earn income and (ii) the shape of the resulting income distribution roughly matches that of the 2018 U.S. economy with trained optimal agent behaviors.
	
	\item RL provides a flexible way to simultaneously optimize and model the behavioral effects of tax policies. RL is instantiated at two levels, that is, for two types of actors: training agent behavioral policy models and a taxation policy model for the social planner. Each actor’s behavioral policy is trained using deep RL, which learns the weights \( \theta_{i} \) of a neural network \( \pi(a_{i, t} | o_{i, t}; \theta_{i}) \) that maps an actor’s observations to actions. Network weights are trained to maximize the expected total discounted reward of the output actions. Specifically, for an agent \( i \) using a behavioral policy \( \pi_{i}(a_{t} | o_{t}; \theta_{i}) \), the RL training objective is (omitting the tax policy \( \pi_{p} \)):
	\begin{equation}\label{Equation3}
		\max_{\pi_{i}} E_{a_{1} \sim \pi_{1}, ..., a_{N} \sim \pi_{N}, s^{'} \sim P} [\sum^{H}_{t = 0} \gamma^{t} r_{t}]
	\end{equation}
	where \( s^{'} \) is the next state and \( P \) denotes the dynamics of the environment. The objective for the planner policy \( \pi_{p} \) is similar. Standard model-free policy gradient methods update the policy weights \( \theta_{i} \) using (variations of):
	\begin{equation}\label{Equation4}
		\mathbf{\triangle \theta_{i}} \propto E_{{a_{1} \sim \pi_{1}, ..., a_{N} \sim \pi_{N}, s^{'} \sim P}}[\sum^{H}_{t = 0} \gamma^{t} r_{t} \nabla_{\theta_{i}} \log \pi_{i}(a_{i, t} | o_{i, t}; \mathbf{\theta_{i}})]
	\end{equation}

	\item In this work, the proximal policy gradients (PPO) is used to train all actors (both agents and planner). To improve learning efficiency, a single-agent policy network \( \pi(a_{i, t} | o_{i, t}; \theta) \) is trained whose weights are shared by all agents, that is, \( \theta_{i} = \theta \). This network is still able to embed diverse, agent-specific behaviors by conditioning on agent-specific observations.
	
	\item At each time step \( t \), each agent observes the following: its nearby spatial surroundings; its current endowment (stone, wood, and coin); private characteristics, such as its building skill; the state of the markets for trading resources; and a description of the current tax rates. These observations form the inputs to the policy network, which uses a combination of convolutional, fully connected, and recurrent layers to represent spatial, non-spatial, and historical information, respectively. For recurrent components, each agent maintains its own hidden state. The policy network for the social planner follows a similar construction but differs somewhat in the information it observes. Specifically, at each time step, the planner policy observes the following: the current inventories of each agent, the state of the resource markets, and a description of the current tax rates. The planner cannot directly observe private information such as an agent’s skill level.
	
	\item Rational economic agents train their policy \( \pi_{i} \) to optimize their total discounted utility over time while experiencing tax rates \( \tau \) set by the planner’s policy \( \pi_{p} \). The agent training objective is:
	\begin{equation}\label{Equation5}
		\forall i : \max_{\pi_{i}} E_{\tau \sim \pi_{p}, a_{i} \sim \pi_{i}, \mathbf{a_{-i}} \sim \mathbf{\pi_{-i}}, s^{'} \sim P} [\sum^{H}_{t = 1} \gamma^{t} r_{i, t} + u_{i, 0}], r_{i, t} = u_{i, t} - u_{i, t - 1}
	\end{equation}
	where the instantaneous reward \( r_{i, t} \) is the marginal utility for agent \( i \) at time step \( t \). Bold-faced quantities denote vectors, and the subscript \( -i \) denotes quantities for all agents except for \( i \).
	
	\item For an agent population with monetary endowments \( \mathbf{C_{t}} = (C_{1, t}, ..., C_{N, t}) \), the equality \( eq(\mathbf{C_{t}}) \) is defined as:
	\begin{equation}\label{Equation6}
		eq(\mathbf{C}_{t}) = 1 - \frac{N}{N - 1} gini(\mathbf{C}_{t}), 0 \leq eq(\mathbf{C}_{t}) \leq 1
	\end{equation}
	where the Gini index is defined as:
	\begin{equation}\label{Equation7}
		gini(\mathbf{C}_{t}) = \frac{\sum_{i = 1}^{N} \sum_{j = 1}^{N} |C_{i, t} - C_{j, t}|}{2N \sum^{N}_{i = 1} C_{i, t}}, 0 \leq gini(\mathbf{C}_{t}) \leq \frac{N - 1}{N}
	\end{equation}

	\item The productivity is defined as the sum of all incomes:
	\begin{equation}\label{Equation8}
		prod(\mathbf{C}_{t}) = \sum_{i} C_{i, t}
	\end{equation}
	The economy is closed: subsidies are always redistributed evenly among agents, and no tax money leaves the system. Hence, the sum of pretax and post-tax incomes is the same. The planner trains its policy \( \pi_{p} \) to optimize social welfare:
	\begin{equation}\label{Equation9}
		\max_{\pi_{p}} E_{\tau \sim \pi_{p}, \mathbf{a} \sim \mathbf{\pi}, s^{'} \sim P} [\sum^{H}_{t = 1} \gamma^{t} r_{p, t} + swf_{0}], r_{p, t} = swf_{t} - swf_{t - 1}
	\end{equation}

	\item The utilitarian social welfare objective is the family of linear-weighted sums of agent utilities, defined for weights \( \omega_{i} \geq 0 \):
	\begin{equation}\label{Equation10}
		swf_{t} = \sum^{N}_{i = 1} \mathbf{\omega}_{i} \cdot \mathbf{u}_{i, t}
	\end{equation}
	Inverse-income is used as the weights: \( \omega_{i} \propto \frac{1}{C_{i}} \), normalized to sum to one. An objective function is defined that optimizes a trade-off between equality and productivity, defined as the product of equality and productivity:
	\begin{equation}\label{Equation11}
		swf_{t} = eq(\mathbf{C}_{t}) \cdot prod(\mathbf{C}_{t})
	\end{equation}
\end{enumerate}

\newpage

\section{Appendix B: The Concordia}
Here is a detailed description of the Concordia framework (\cite{Vezhnevets2023}):

\begin{enumerate}
	\item A generative modelling of social interactions have two parts: the model of the environment and the model of individual behaviour. In Concordia both are generative. Thus in Concordia there are : (1) generative agents and (2) a generative model for the environment, space, or world where the social interactions take place. In Concordia, the model which is responsible for the environment is called the game master. The game master name and the way Concordia works were inspired by table-top role-playing games like Dungeons and Dragons where a game master takes the role of the storyteller. In these games, players interact with one another in a hypothetical world invented by the game master.
	
	\item Concordia agents receive observations of the environment as inputs and based on those generate actions. The game master receives the agent actions and produces event statements, which define the course of events in the simulation as a result of the agent’s generated action. The game master also produces and sends observations to agents. Observations, actions, and event statements are all strings in a natural language like English. The game master is also responsible for keeping up-to-date the grounded variables, moving forward the clock, and advancing the episode loop.
	
	\item Concordia agents generate their behaviours by explaining what they want to do in natural language. The game master absorbs their intended actions, decides on the outcome of their attempts, and generates event statements. Basically, the game master is performing the following things: (1) keeping a grounded and consistent state of the world where agents interact with each other, (2) communicating the observable state of the world to the agents, (3) deciding the effect of agents’ actions on the world and each other, (4) resolving the issues when actions submitted by multiple agents conflict with one another.
	
	\item The game master’s most important responsibility is to provide the grounding for particular experimental variables, which are defined for an experiment. The game master determines the effect of the agents’ actions on these variables, records them, and checks that they are valid. Whenever an agent tries to perform an action that somehow contradicts the grounding, it conveys to them that its action is invalid. For example, in an economic simulation the amount of money in an agent’s possession may be a grounded variable. The game master would track whether agents gained or lost money on each step and perhaps prevent them from paying more than they have available.
	
	\item The produced agents behaviours should be consistent with common sense, in accordance by social norms, and individually grounded based on a personal history of past events as well as ongoing understanding of the current situation.
	
	\item It is argued that humans generally act as though they choose their actions by answering three key questions: (1) What kind of situation is this? (2) What kind of person am I? (3) What does a person such as I do in a situation such as this?
	
	\item The premise behind Concordia is that since modern LLMs have been trained on huge amounts of human culture, they are thus capable of giving reasonable answers to the above questions when provided with the historical context of a particular agent. The idea is that, if the outputs of LLMs conditioned to model specific human populations, they reflect the beliefs and attitudes of those populations. It is also shown that personality measurements in the outputs of some LLMs under specific prompting configurations are trustable, therefore generative agents could be used to model humans with diverse psychological profiles. In some cases answering the key questions might require common sense reasoning and/or planning, which LLMs show capacity for and show similar biases in behavioural economics experiments as humans. The ability of LLMs to learn ‘in-context’ and ‘zero-shot’ reinforces the hypothesis further that the agent might be able to understand what is expected of them in the current situation from a demonstration.
	
	\item The next step is to make available a record of an agent’s historical experience to an LLM so it would be able to answer the above mentioned key questions. Concordia makes this possible by using an associative memory in a modular and flexible fashion to keep the record of agents experience.
	
	\item Memory is a set of strings \( \textbf{m} \) that records everything remembered or currently experienced by the agent. The working memory is \( {z^{i}}_{i} \) composed of the states of individual components. A component \( i \) has a state \( z^{i} \), which is statement in natural language. The components update their states by querying the memory containing the incoming observations, and using LLM for summarising and reasoning. Components can also condition their update on the current state of other components. As an instance, the planning component can update its state if an incoming observation invalidates the current plan, conditioned on the state of the 'goal’ component.
	
	\item The incoming observations are fed into the agents memory to make them available when components update. When creating a generative agent in Concordia, the user creates the components that are relevant for their simulations. They decide on the initial state and the update function. The components are then supplied to the agents constructor. Formally, the agent is defined as a two step sampling process, using a LLM \( P \). In the action step, the agent samples its activity \( a_{t} \), given the state of components \( \textbf{z}_{t} = {z_{t}^{i}}_{i} \):
	\begin{equation}\label{Equation12}
		a_{t} \sim p(. \mid f^{a}(z_{t}))
	\end{equation}

    \item Here \( f_{a} \) is a formatting function, which out of the states of components creates the grounding used to sample the action to take. The most simple form of \( f_{a} \) is a concatenation operator over \( \textbf{z}_{t} = {z_{t}^{i}}_{i} \). We do not explicitly condition on the memory \( \textbf{m} \) or observation \( \textbf{o} \), since we can subsume them into components. First, we can immediately add \( \textbf{o}_{t} \) to the memory \( \textbf{m}_{t} = \textbf{m}_{t-1} \cup \textbf{o}_{t} \). Unlike RL, we do not assume that the agent responds with an action to every observation. The agent can get several observations before it acts, therefore \( \textbf{o}_{t} \) is a set of strings. Then we can set \( \textbf{z}_{0} \) to be the component that incorporates the latest observations and relevant memories into its state. This allows us to exclusively use the components to define the agent. In the second step, the agent samples its state \( \textbf{z} \), given the agents memory \( \textbf{m}_{t} \) up to the present time: 
    \begin{equation}\label{Equation13}
    	z^{i}_{t+1} \sim p(. \mid f^{i}(\textbf{z}_{t}, \textbf{m}_{t}))
    \end{equation}

    \item Here, \( f^{i} \) is a formatting function that turns the memory stream and the current state of the components into the query for the component update. In Concordia, conditioning is explicitely done on the memory stream \( \textbf{m} \), since a component may make specific queries into the agent’s memory to update its state. Here Eq.~\ref{Equation13} updates components after every action, but generally, it is up to the agent to decide at which step to update each of its components. It is reasonable to update some components less frequently for efficiency or longer term consistency.
    
    \item The game master takes care of all aspects of the simulated world not directly controlled by the agents. The game master mediates between the state of the world and agents’ actions. The state of the world is contained in game master’s memory and the values of grounded variables such as inventories or votes. To achieve this the game master has to repeatedly answer the following questions: (1) What is the state of the world? (2) Given the state of the world, what event is the outcome of the players activity? (3) What observation do players make of the event? (4) What effect does the event have on grounded variables?
    
    \item The game master is implemented in a similar fashion to a generative agent. Like agents, the game master has an associative memory implemented using various components. However, instead of contextualising action selection, the components of the game master describe the state of the world, for example location and status of players, state of grounded variables (inventories, votes) and so on.  Thus the game master can decide the event that happens as the outcome of players’ actions. The outcome is described in the event statement which is then added to the game master associative memory. After the event has been decided the game master elaborates on its consequences. For example, the event could have changed the value of one of the grounded variables.
    
    \item The game master generates an event statement  \( e_{t} \) in response to each agent action:
    \begin{equation}\label{Equation14}
    	e_{t} \sim p(. \mid f^{e}(\textbf{z}_{t}, {a}_{t}))
    \end{equation}

    \item The above equation highlights the fact that the game master generates an event statement \( e_{t} \) in response to every action of any agent, while the agent might take in several observations before it acts (or none at all). After adding the event statement \( e_{t} \) to its memory the game master can update its components using the same Eq.~\ref{Equation13} as the agent. It can then emit observations \( o^{i}_{t} \) for player \( i \) using the following equation:
    \begin{equation}\label{Equation15}
    	\textbf{O}^{i}_{t+1} \sim p(. \mid f^{o}(\textbf{z}_{t+1}))
    \end{equation}

    \item In case the game master judges that a player did not observe the event, no observation is emitted. Notice that the components can have their internal logic written using any existing modelling tools (ODE, graphical models, finite state machines, etc.) and therefore can bring known models of certain physical, chemical, or financial phenomena into the simulation.
    
\end{enumerate}

\newpage

\section{Appendix C: Supplemental Figures}

\begin{figure}[h]
	\centering
	\includegraphics[width=0.7\linewidth]{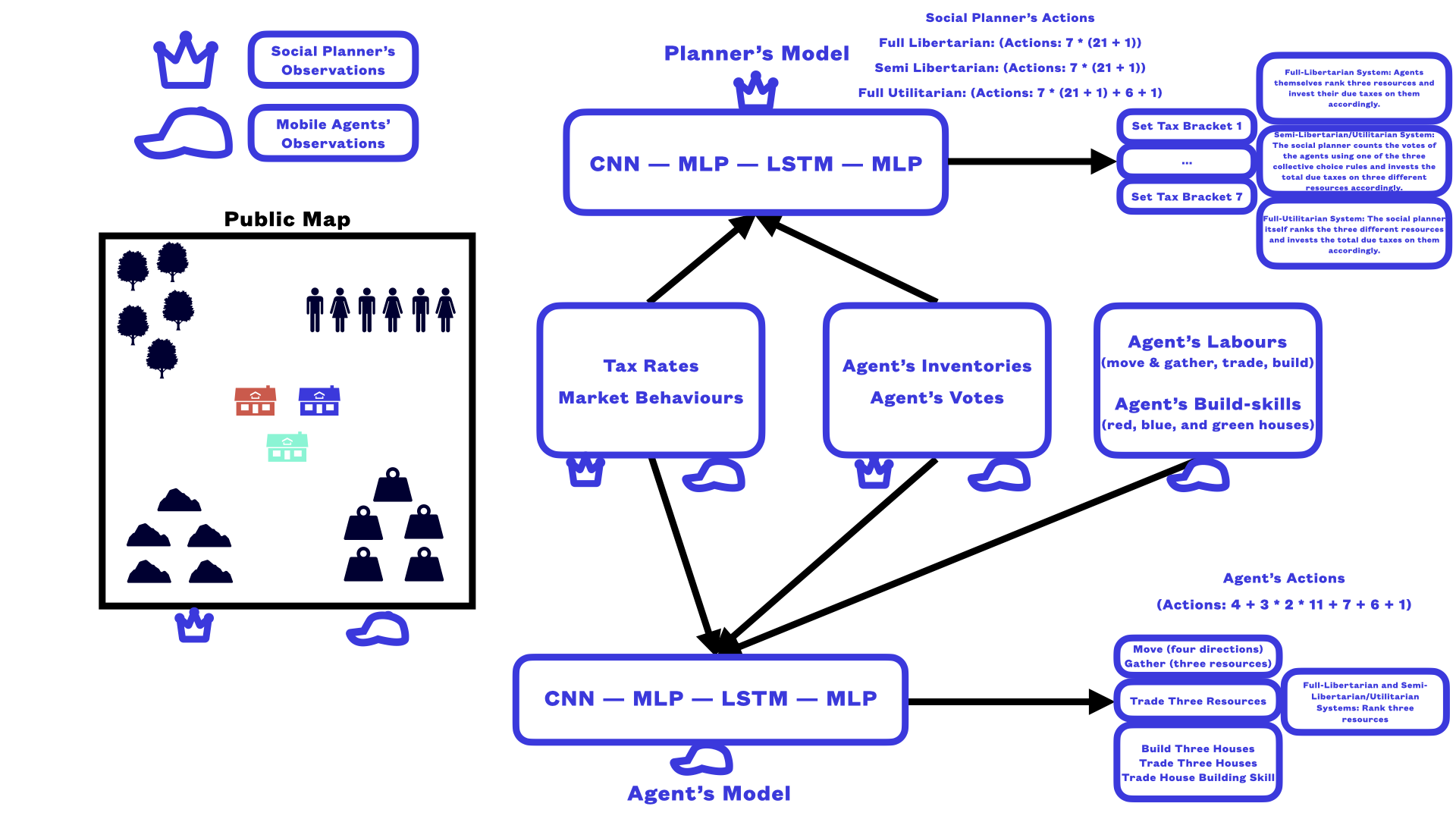}
	\caption{Observation and action spaces for economic mobile agents and the central social planner. The agents and the planner observe different subsets of the world state. Agents observe the public spatial map, tax rates, market prices, inventories, votes, labours, and skill levels. Agents can decide to move (and therefore gather if moving onto a resource), buy, sell, build, trade, or vote. There are maximum 84 unique actions available to the agents. The planner observes the public spatial map, tax rates, market prices, agent inventories, and votes. The planner decides how to set tax rates, choosing one of 22 settings for each of the 7 tax brackets. It has maximum 161 unique actions. MLP: multi-layer perceptron, LSTM: long short-term memory, CNN: convolutional neural network. This figure should be compared to Fig. 9 of the original AI-Economist paper (\cite{Zheng2022}).}
	\label{Figure13}
\end{figure}

\newpage

\begin{figure}
	\centering
	\includegraphics[width=0.7\linewidth]{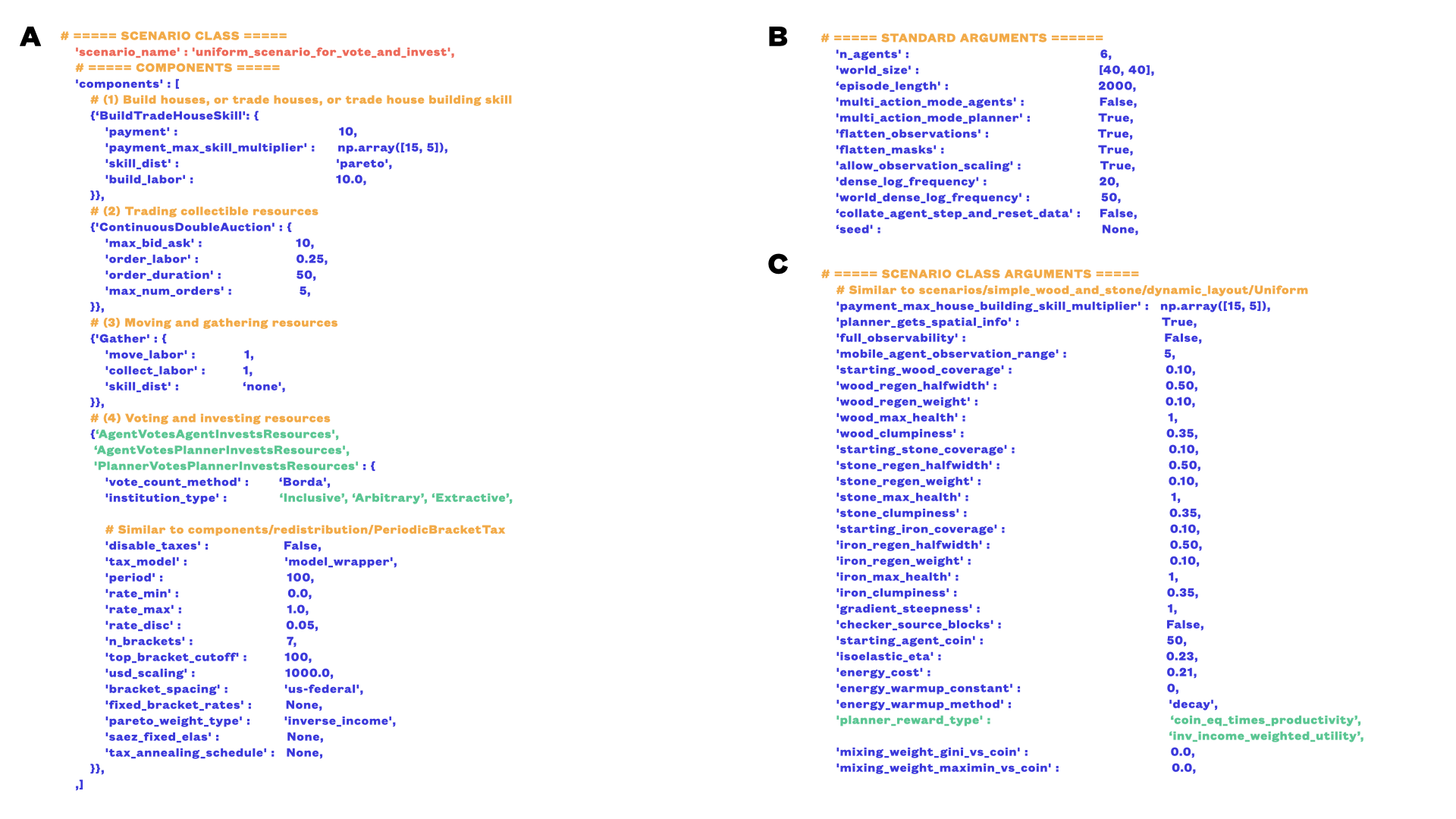}
	\caption{(A)(B)(C) Different features and input parameters of the extended AI-Economist which are used and their aggregated plots are brought and discussed in the main text. The orange texts indicate various parts of the input structure. The green texts show the alternative parameters which are tested in this paper.}
	\label{Figure14}
\end{figure}

\newpage

\begin{figure}
	\centering
	\includegraphics[width=0.7\linewidth]{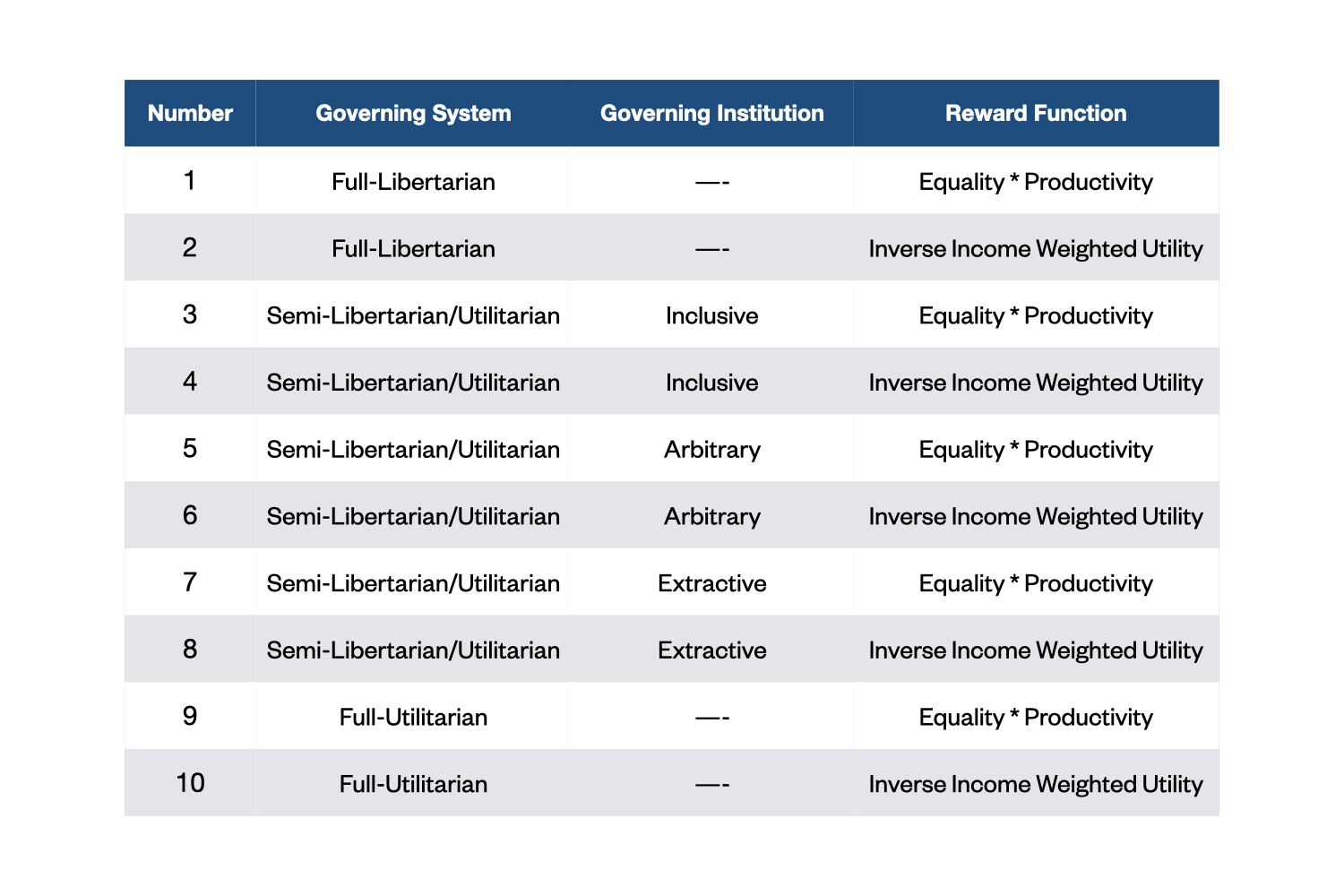}
	\caption{A table showing all different runs of the extended AI-Economist using different values as input parameters. The \textit{Governing System} is an entity along individualistic-collectivistic axis, while \textit{Governing Institution} is an entity along discriminative axis. The \textit{Reward Function} refers to the reward function of the central planner. To generate the plots in the main text, the generated results of a pair of consecutive somewhat similar simulations are pooled together.}
	\label{Figure15}
\end{figure}

\newpage

\begin{figure}
	\centering
	\includegraphics[width=0.7\linewidth]{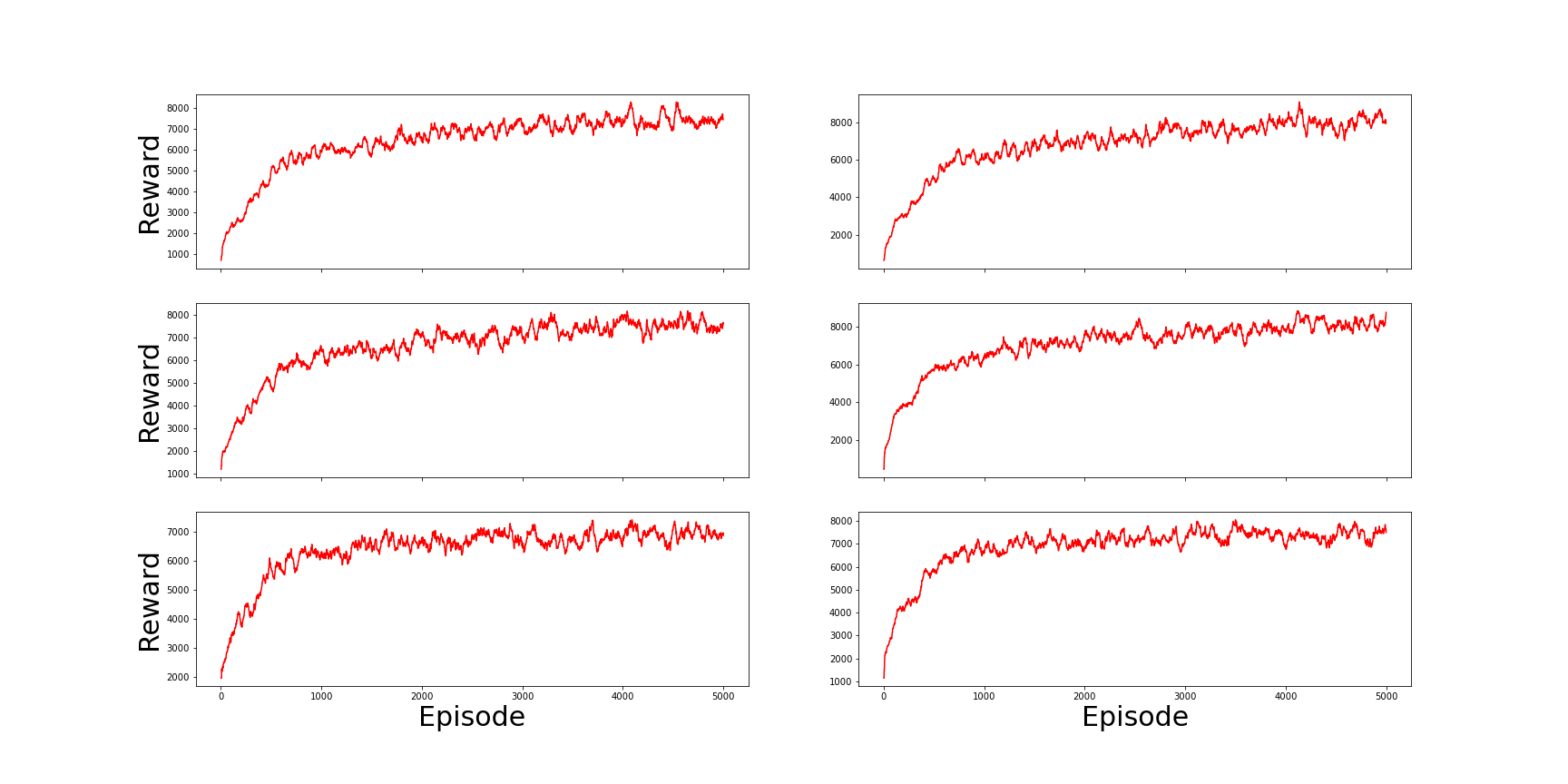}
	\caption{Average episode reward across training (5000 episodes) for all runs of the extended AI-Economist along individualistic-collectivistic axis (Fig.~\ref{Figure15}). The first row shows the relevant plots of the two runs of Full-Libertarian governing system. Moreover, the second row shows the plots of the two runs of Semi-Libertarian/Utilitarian governing system with Inclusive governing institution. Finally, the third row shows the plots of the two runs of Full-Utilitarian governing system. It is worthwhile to mention that the training of two-level RL is particularly unstable, but it seems that almost all the simulations have been converged.}
	\label{Figure16}
\end{figure}

\newpage

\begin{figure}
	\centering
	\includegraphics[width=0.7\linewidth]{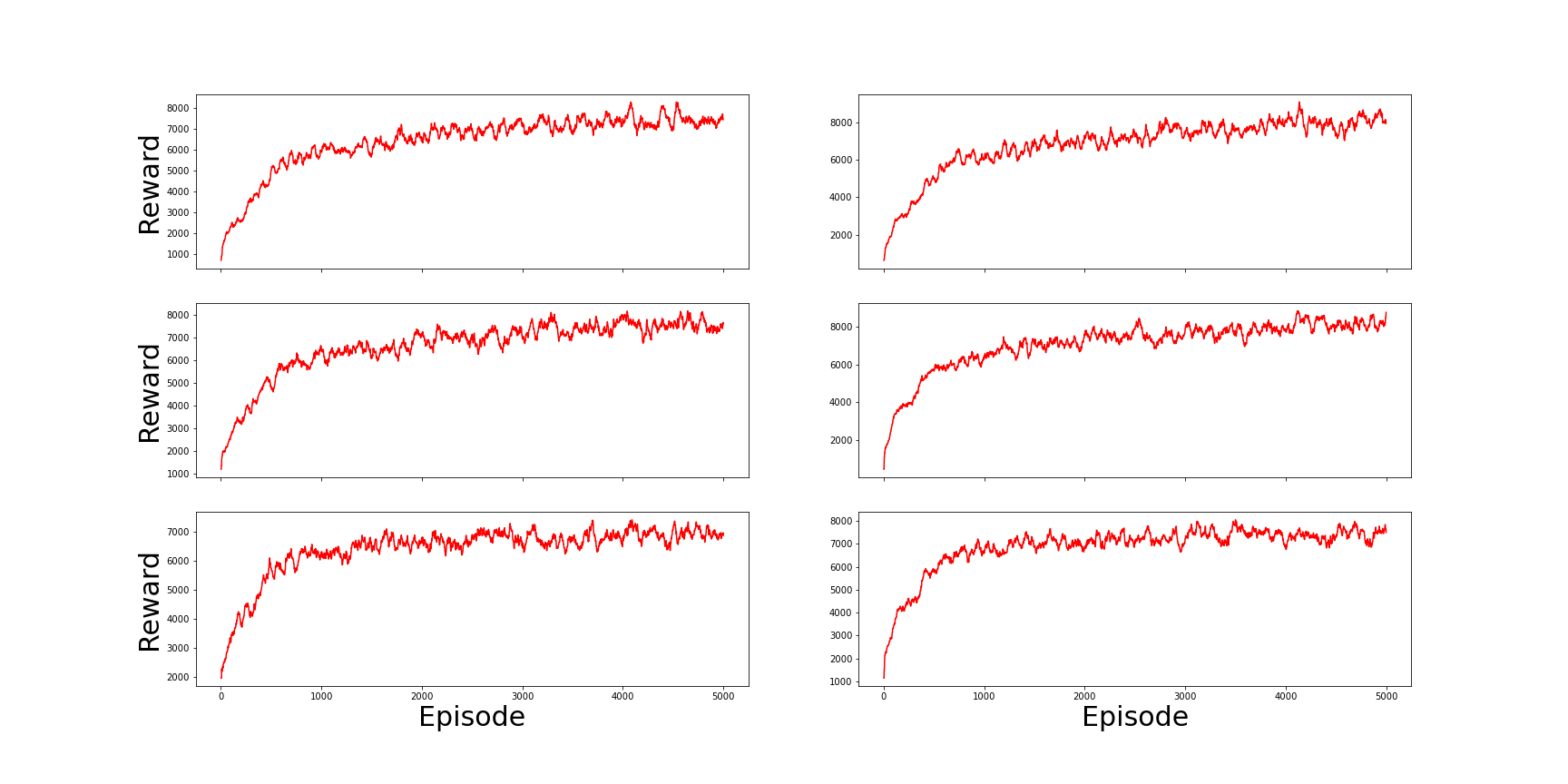}
	\caption{Average episode reward across training (5000 episodes) for all runs of the extended AI-Economist along discriminative axis (Fig.~\ref{Figure15}). The first row shows the relevant plots of the two runs of Inclusive governing institution of the Semi-Libertarian/Utilitarian governing system. Moreover, the second row shows the plots of the two runs of Arbitrary governing institution of the Semi-Libertarian/Utilitarian governing system. Finally, the third row shows the plots of the two runs of Extractive governing institution of the Semi-Libertarian/Utilitarian governing system. It is worthwhile to mention that the training of two-level RL is particularly unstable, but it seems that almost all the simulations have been converged.}
	\label{Figure17}
\end{figure}

\newpage

\begin{figure}
	\centering
	\includegraphics[width=0.7\linewidth]{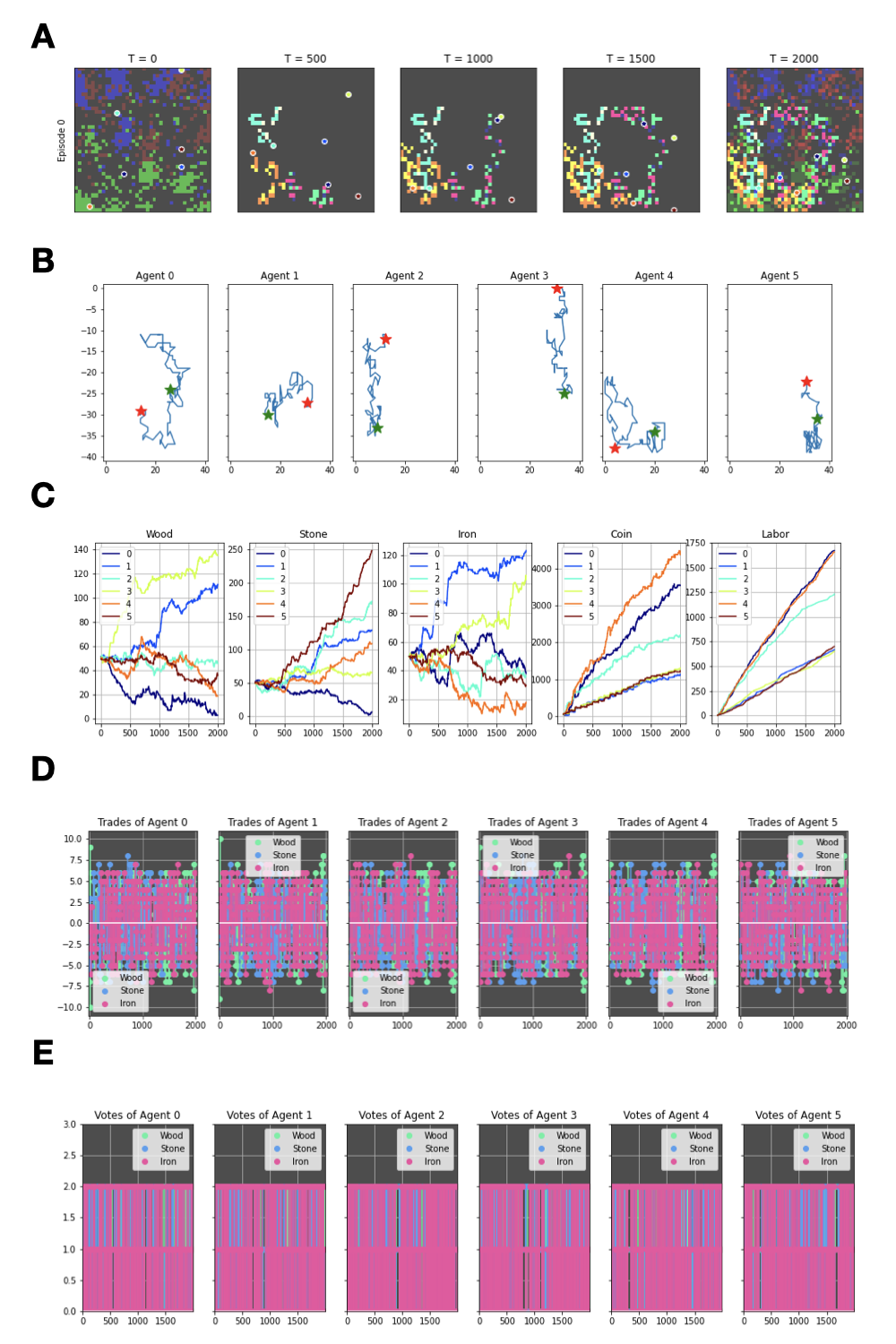}
	\caption{Sample plots obtained from running the extended AI-Economist under Full-Libertarian governing system with equality times productivity as the objective function of the central planner. (A) The environment across five time-points of an episode, (B) the movement of the agents across an episode, (C) the budgets of three resources plus coin and labour of the agents across an episode, (D) the trades of three resources of the agents across an episode, (E) and the votes of the agents across an episode.}
	\label{Figure18}
\end{figure}

\newpage

\begin{figure}
	\centering
	\includegraphics[width=0.7\linewidth]{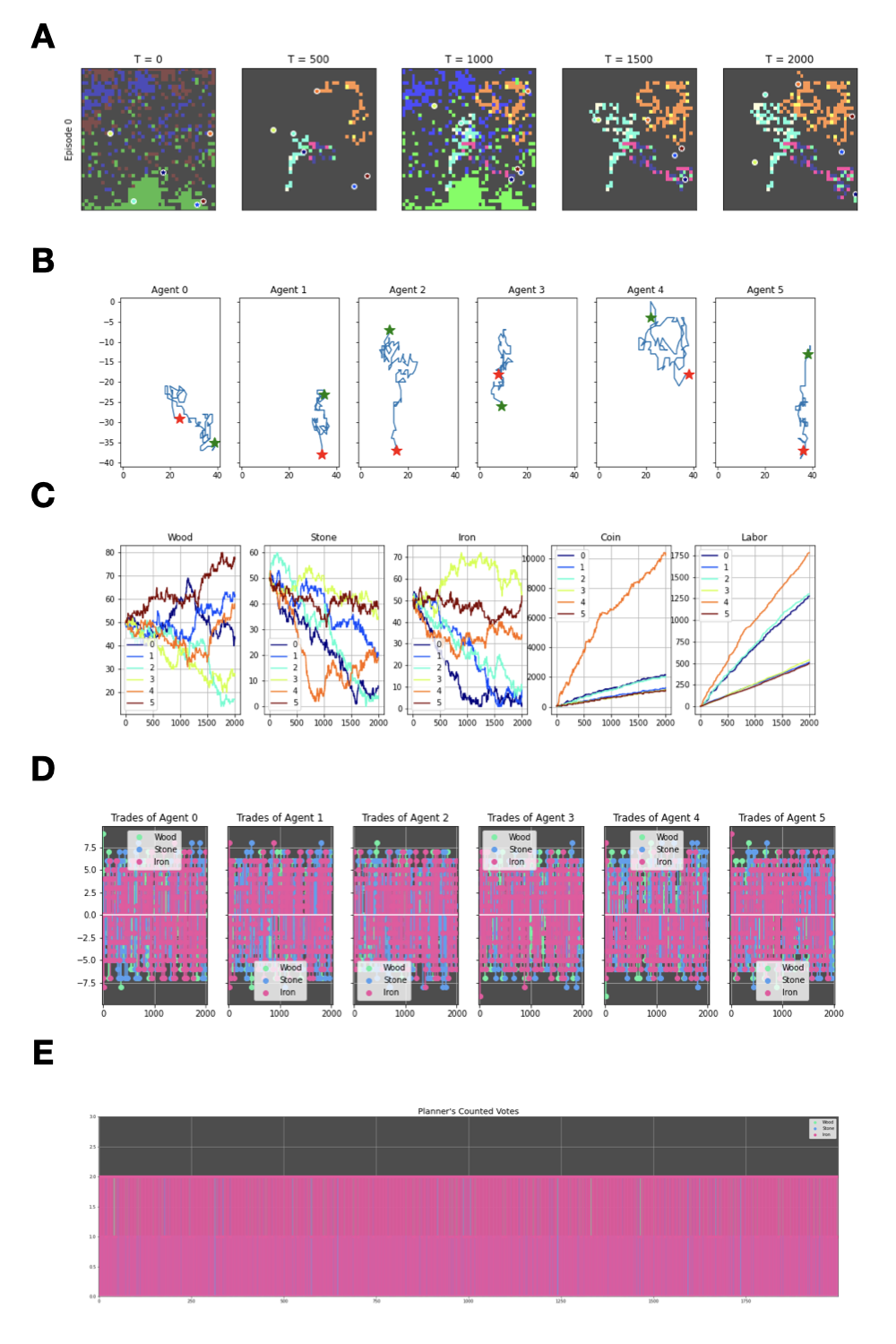}
	\caption{Sample plots obtained from running the extended AI-Economist under Semi-Libertarian/Utilitarian governing system with equality times productivity as the objective function of the central planner. (A) The environment across five time-points of an episode, (B) the movement of the agents across an episode, (C) the budgets of three resources plus coin and labour of the agents across an episode, (D) the trades of three resources of the agents across an episode, (E) and the counted votes of the agents across an episode.}
	\label{Figure19}
\end{figure}

\newpage

\begin{figure}
	\centering
	\includegraphics[width=0.7\linewidth]{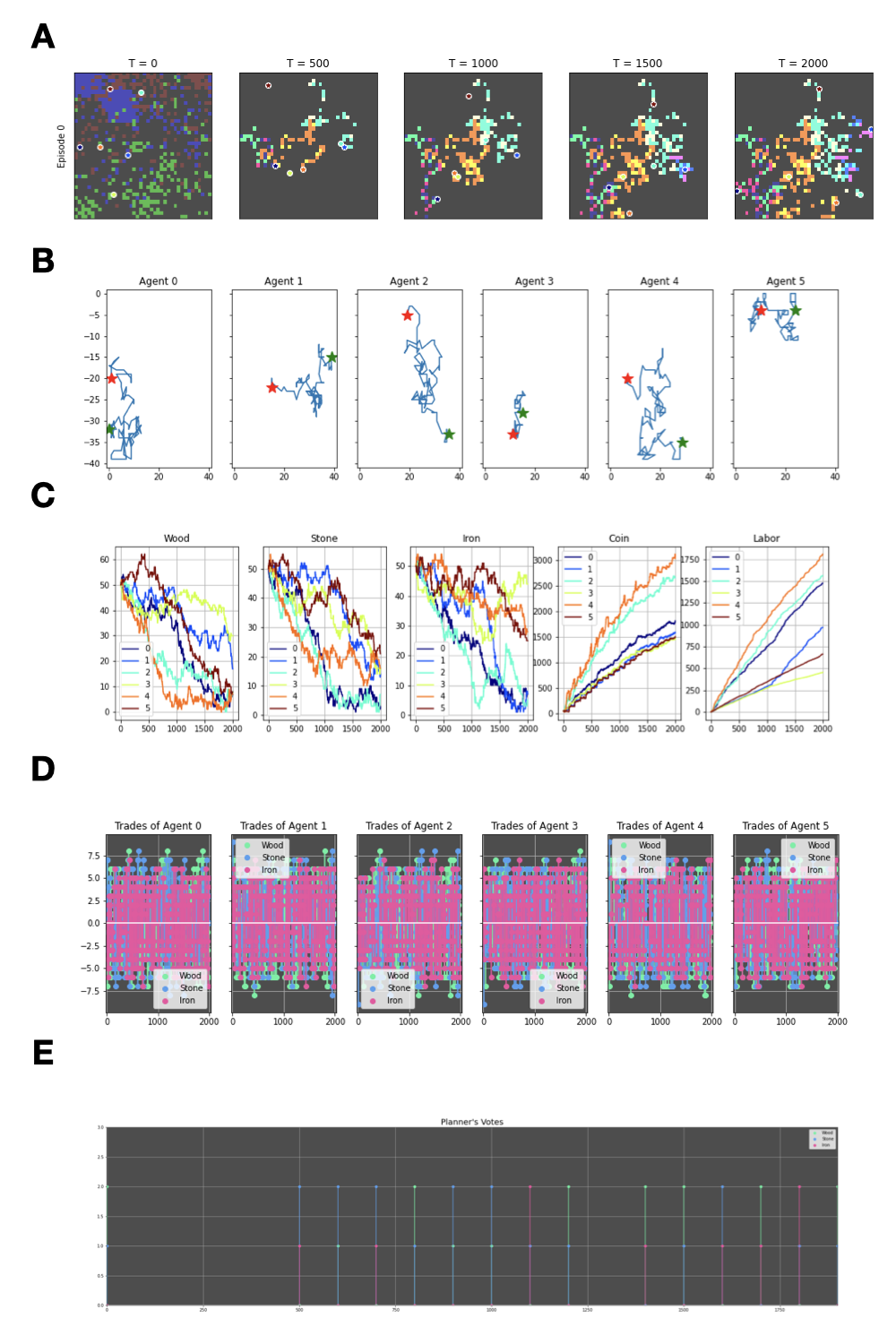}
	\caption{Sample plots obtained from running the extended AI-Economist under Full-Utilitarian governing system with equality times productivity as the objective function of the central planner. (A) The environment across five time-points of an episode, (B) the movement of the agents across an episode, (C) the budgets of three resources plus coin and labour of the agents across an episode, (D) the trades of three resources of the agents across an episode, (E) and the votes of the central planner across an episode.}
	\label{Figure20}
\end{figure}

\newpage

\begin{figure}
	\centering
	\includegraphics[width=0.7\linewidth]{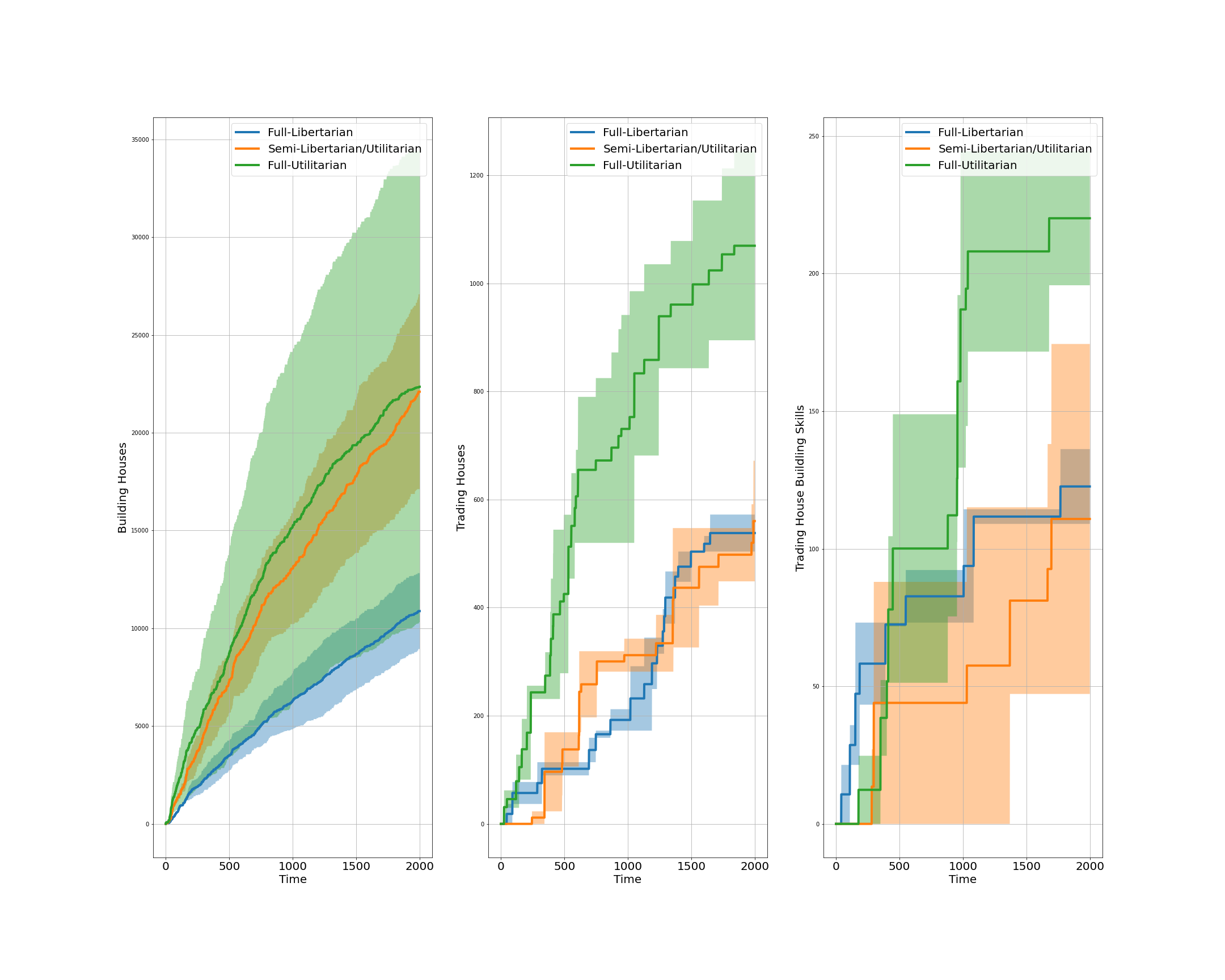}
	\caption{Building houses, trading houses, and trading house building skill across an episode for three governing systems of the extended AI-Economist: Full-Libertarian, Semi-Libertarian/Utilitarian, and Full-Utilitarian. These are obtained by averaging over two similar simulations per governing system differing only in the type of the reward function of the central planner (Fig.~\ref{Figure15}). This figure should be compared to (Fig.~\ref{Figure4}) in the main text, which I believe it is more informative than this figure.}
	\label{Figure21}
\end{figure}

\newpage

\begin{figure}
	\centering
	\includegraphics[width=0.7\linewidth]{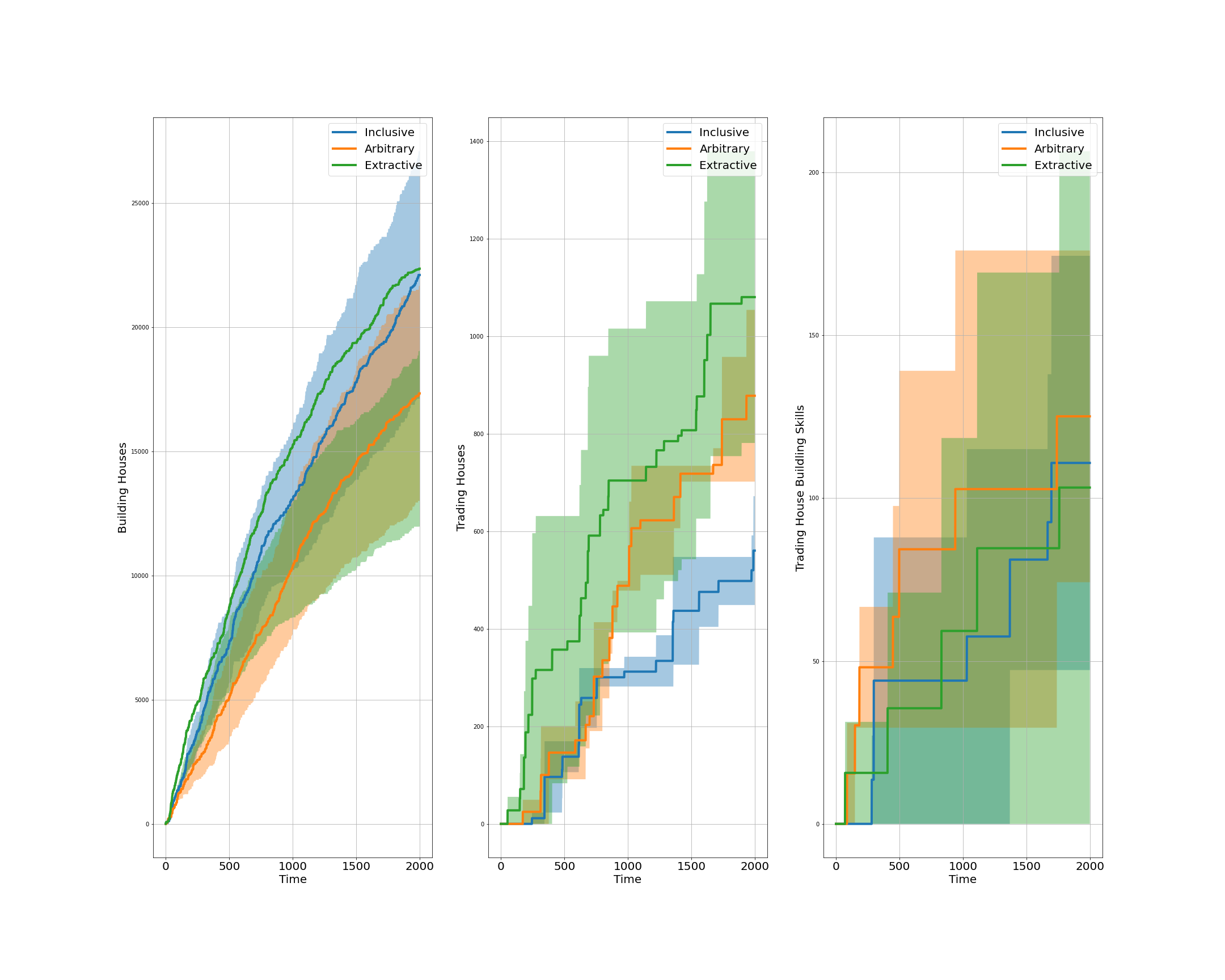}
	\caption{Building houses, trading houses, and trading house building skill across an episode for three governing institutions of the Semi-Libertarian/Utilitarian governing system of the extended AI-Economist: Inclusive, Arbitrary, and Extractive. These are obtained by averaging over two similar simulations per governing institution differing only in the type of the reward function of the central planner (Fig.~\ref{Figure15}). This figure should be compared to (Fig.~\ref{Figure5}) in the main text, which I believe it is more informative than this figure.}
	\label{Figure22}
\end{figure}

\newpage

\begin{figure}
	\centering
	\includegraphics[width=0.7\linewidth]{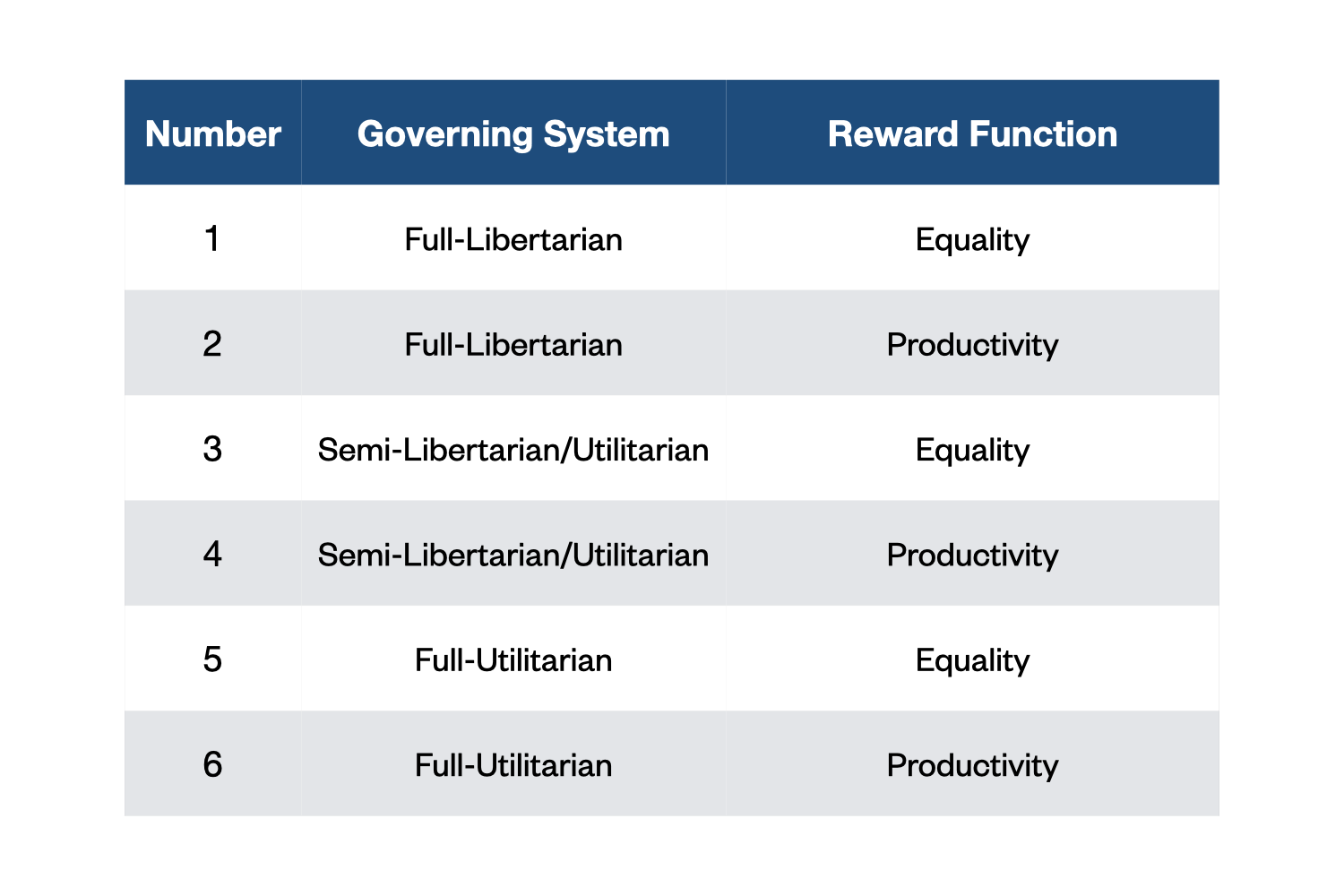}
	\caption{A table showing all different runs of the extended Concordia using different values as input parameters. The \textit{Reward Function} refers to the social reward function of the central planner or the game master.}
	\label{Figure23}
\end{figure}

\newpage

\begin{figure}
	\centering
	\includegraphics[width=0.7\linewidth]{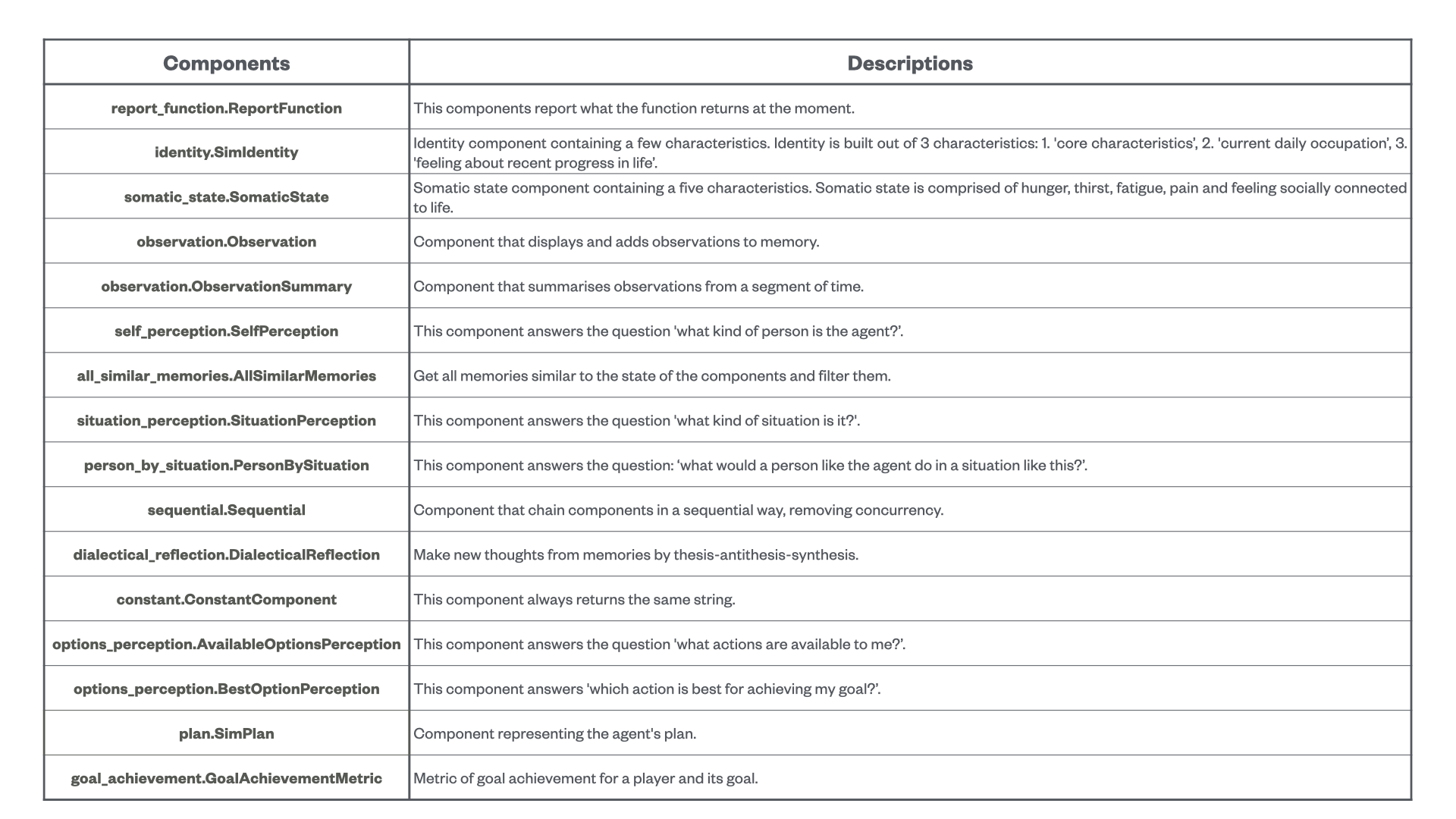}
	\caption{Different components used in the architecture of the agents or players of the extended Concordia and their descriptions.}
	\label{Figure24}
\end{figure}

\newpage

\begin{figure}
	\centering
	\includegraphics[width=0.7\linewidth]{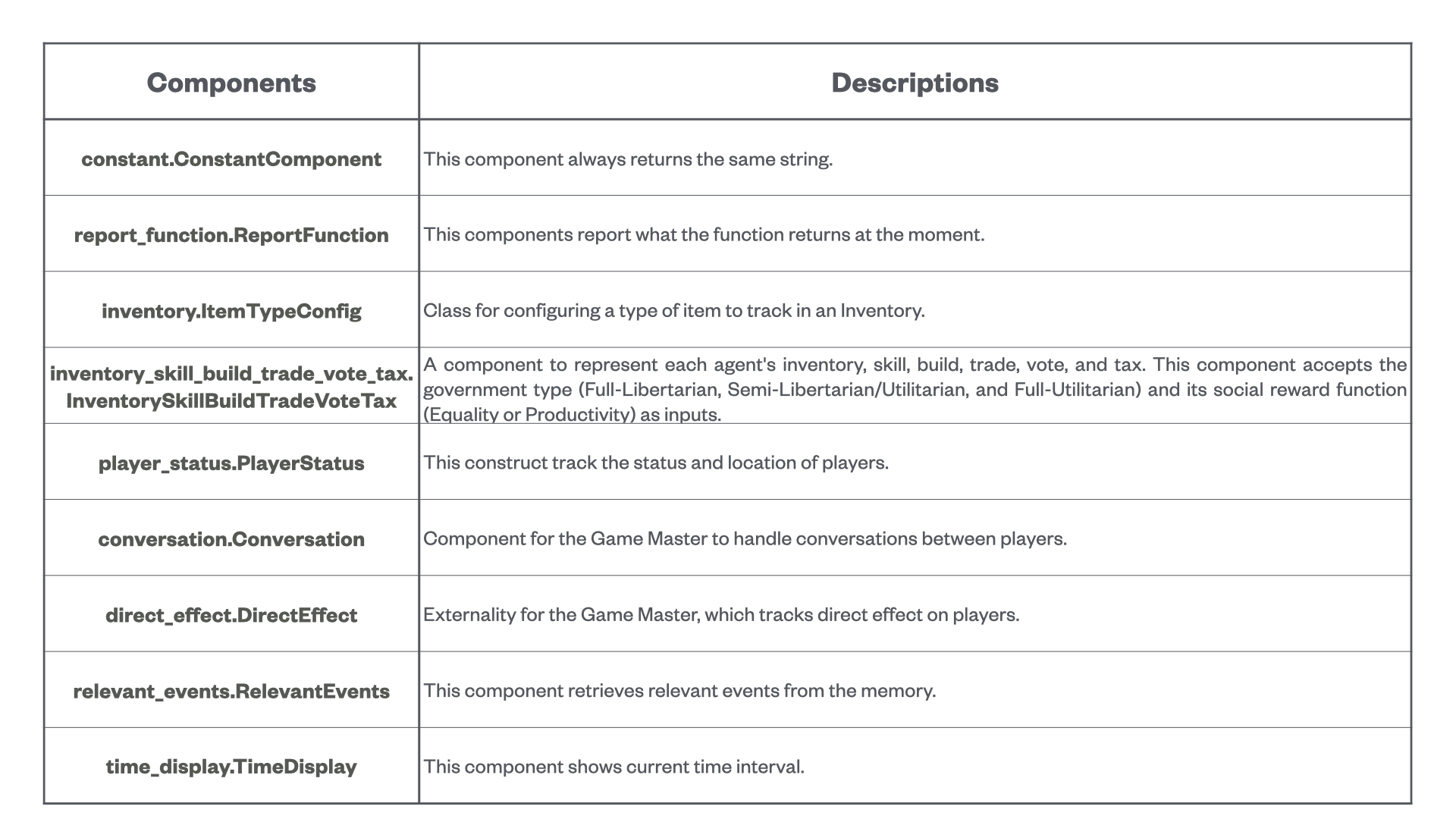}
	\caption{Different components used in the architecture of the game master of the extended Concordia and their descriptions.}
	\label{Figure25}
\end{figure}

\newpage

\begin{figure}
	\centering
	\includegraphics[width=0.7\linewidth]{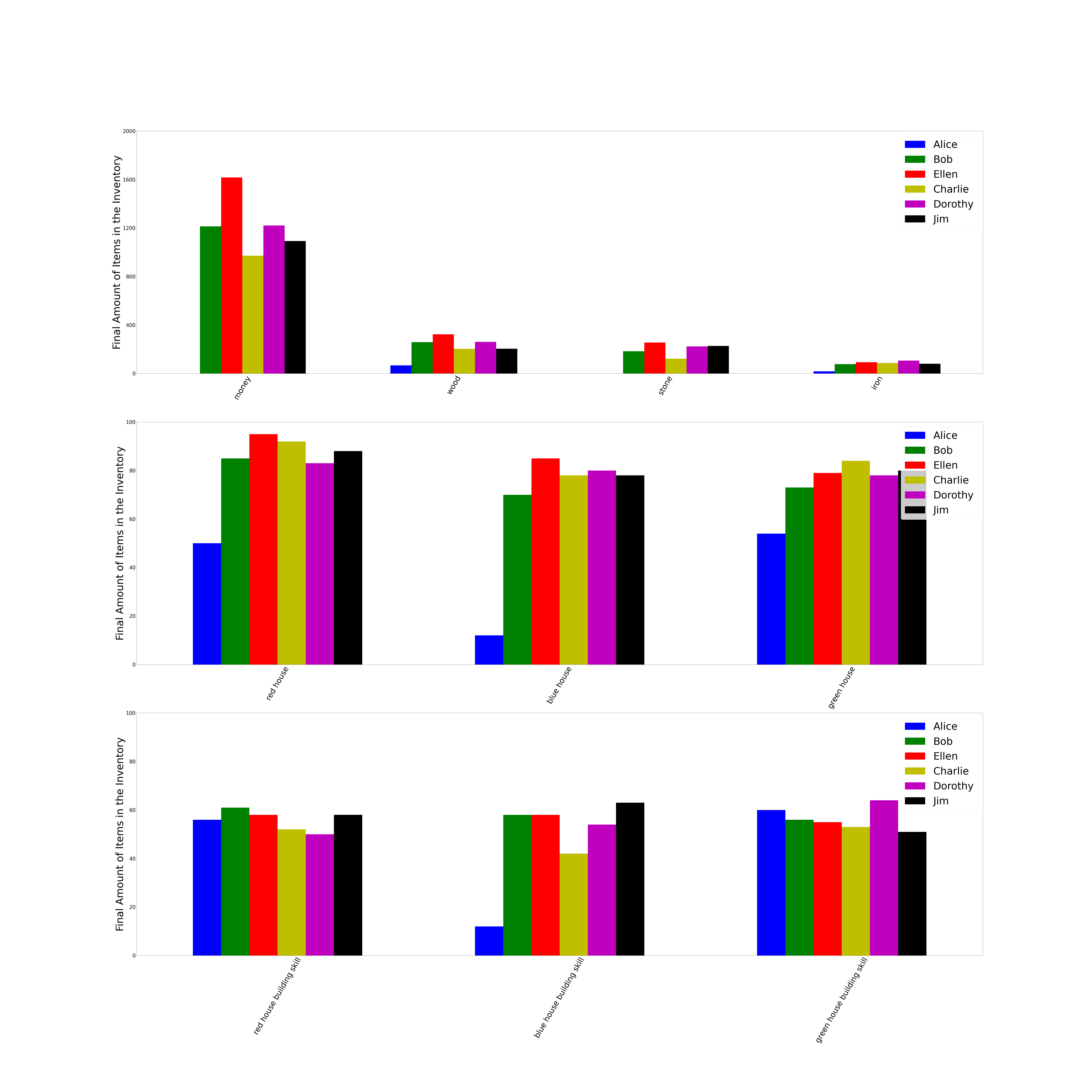}
	\caption{Final amounts of items in the inventories of six agents of the extended Concordia under Full-Libertarian governing system when the game master cares about equality in the society.}
	\label{Figure26}
\end{figure}

\newpage

\begin{figure}
	\centering
	\includegraphics[width=0.7\linewidth]{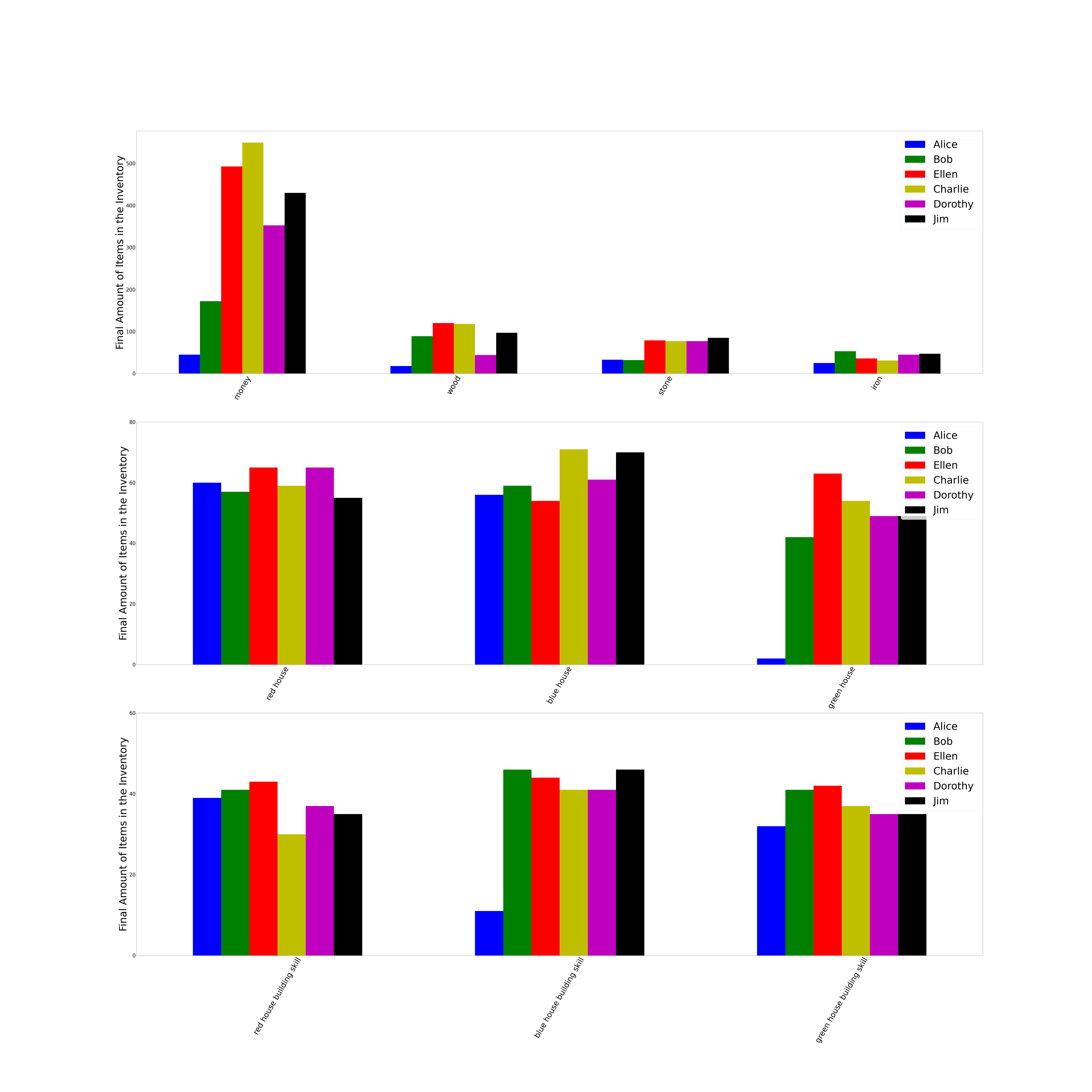}
	\caption{Final amounts of items in the inventories of six agents of the extended Concordia under Full-Libertarian governing system when the game master cares about productivity in the society.}
	\label{Figure27}
\end{figure}

\newpage

\begin{figure}
	\centering
	\includegraphics[width=0.7\linewidth]{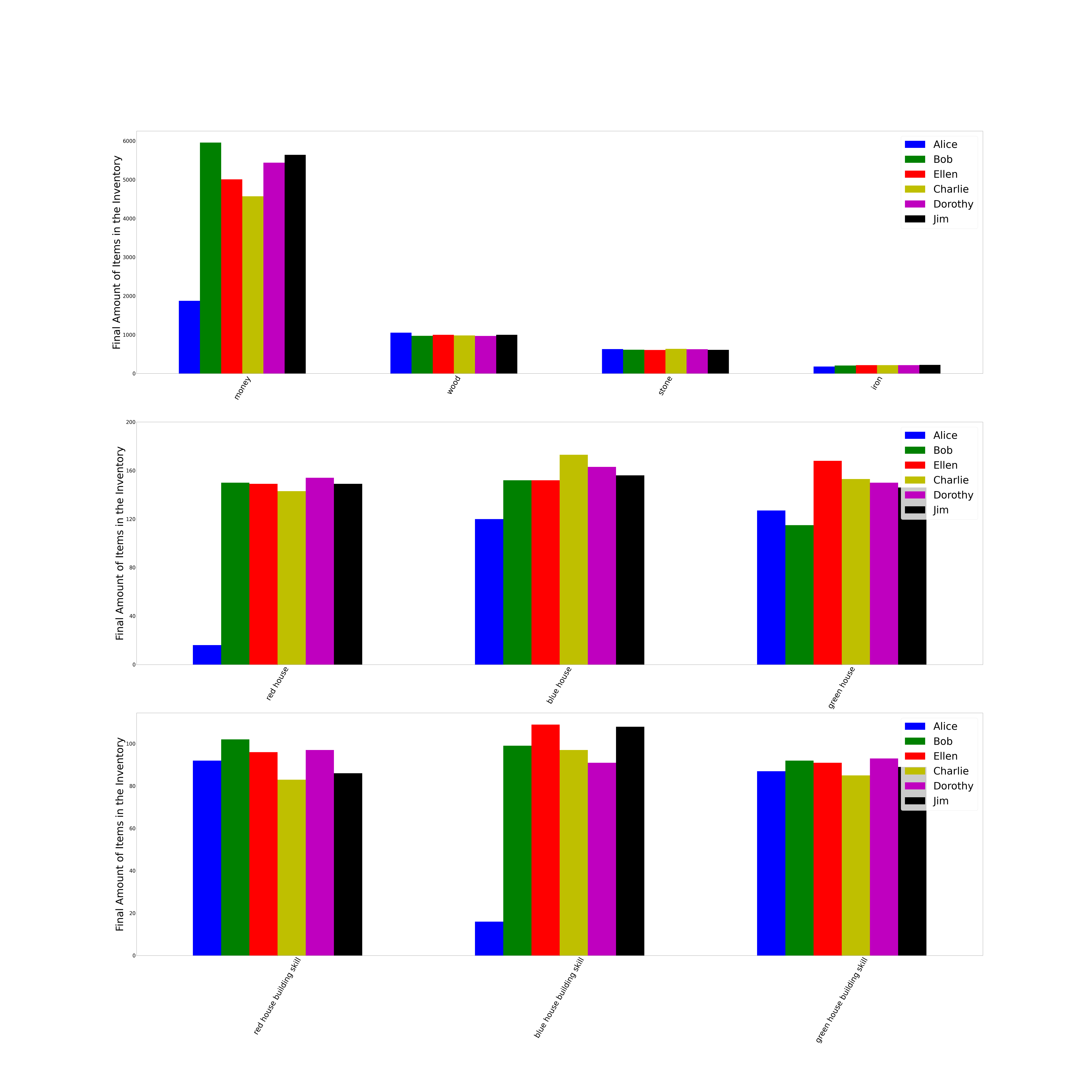}
	\caption{Final amounts of items in the inventories of six agents of the extended Concordia under Semi-Libertarian/Utilitarian governing system when the game master cares about equality in the society.}
	\label{Figure28}
\end{figure}

\newpage

\begin{figure}
	\centering
	\includegraphics[width=0.7\linewidth]{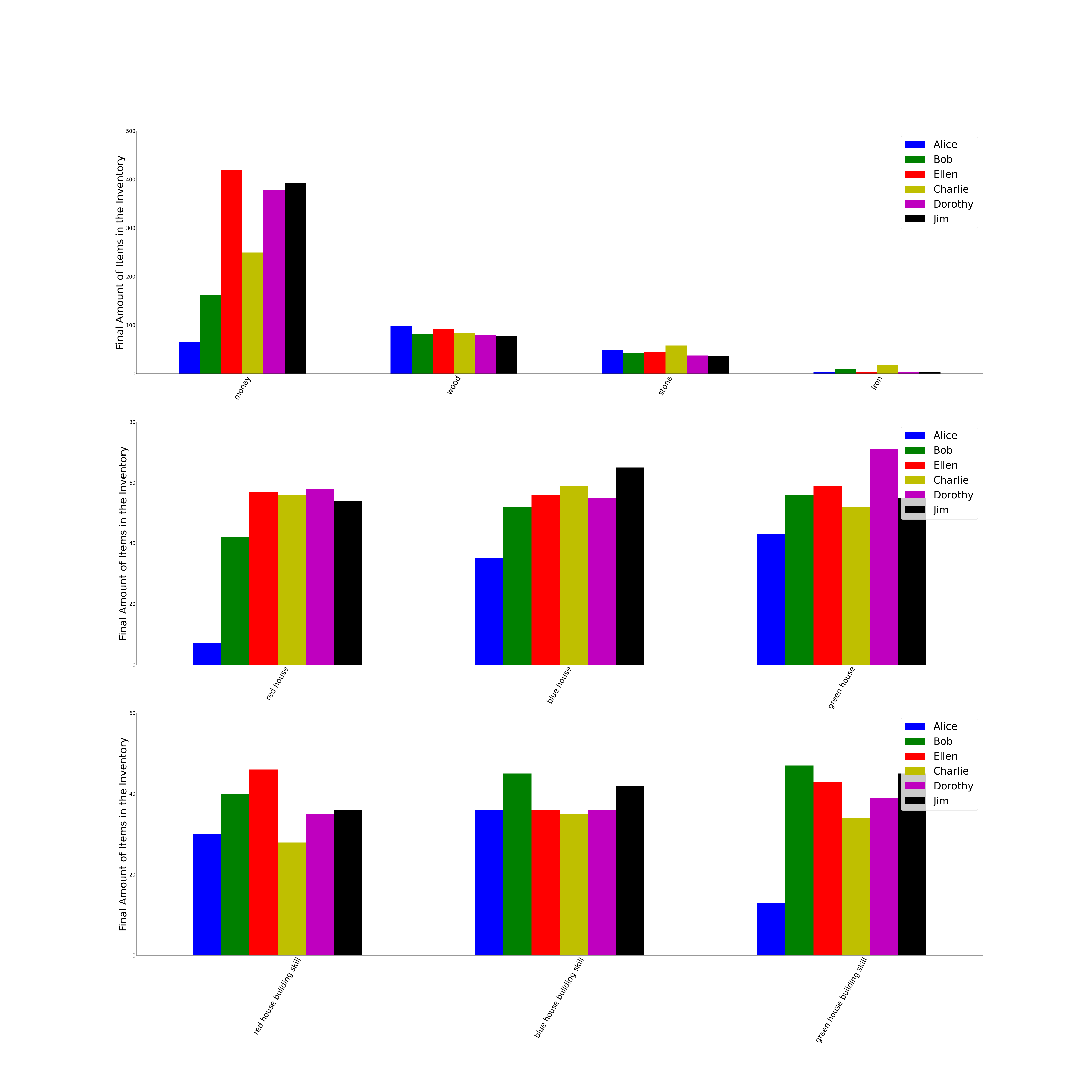}
	\caption{Final amounts of items in the inventories of six agents of the extended Concordia under Semi-Libertarian/Utilitarian governing system when the game master cares about productivity in the society.}
	\label{Figure29}
\end{figure}

\newpage

\begin{figure}
	\centering
	\includegraphics[width=0.7\linewidth]{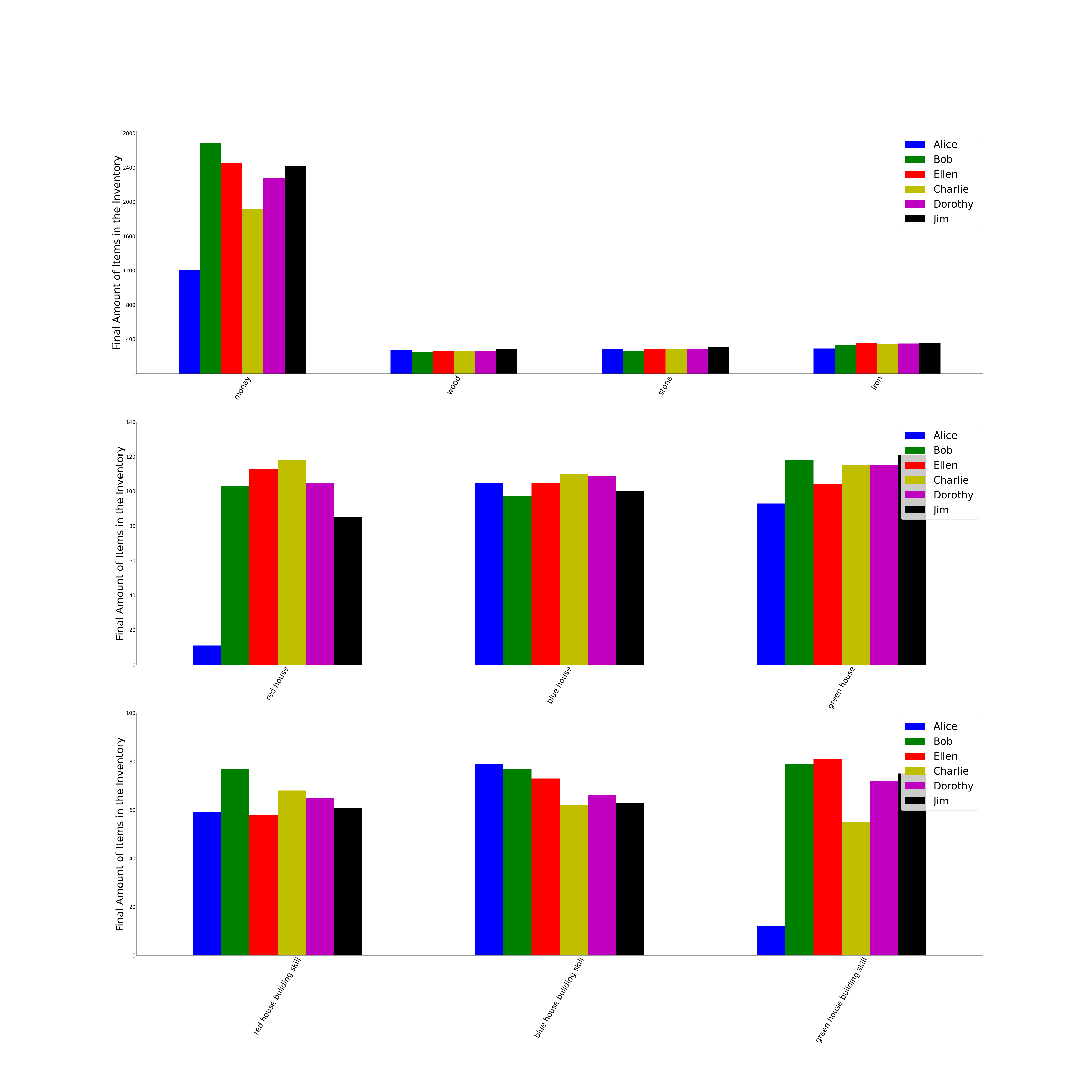}
	\caption{Final amounts of items in the inventories of six agents of the extended Concordia under Full-Utilitarian governing system when the game master cares about equality in the society.}
	\label{Figure30}
\end{figure}

\newpage

\begin{figure}
	\centering
	\includegraphics[width=0.7\linewidth]{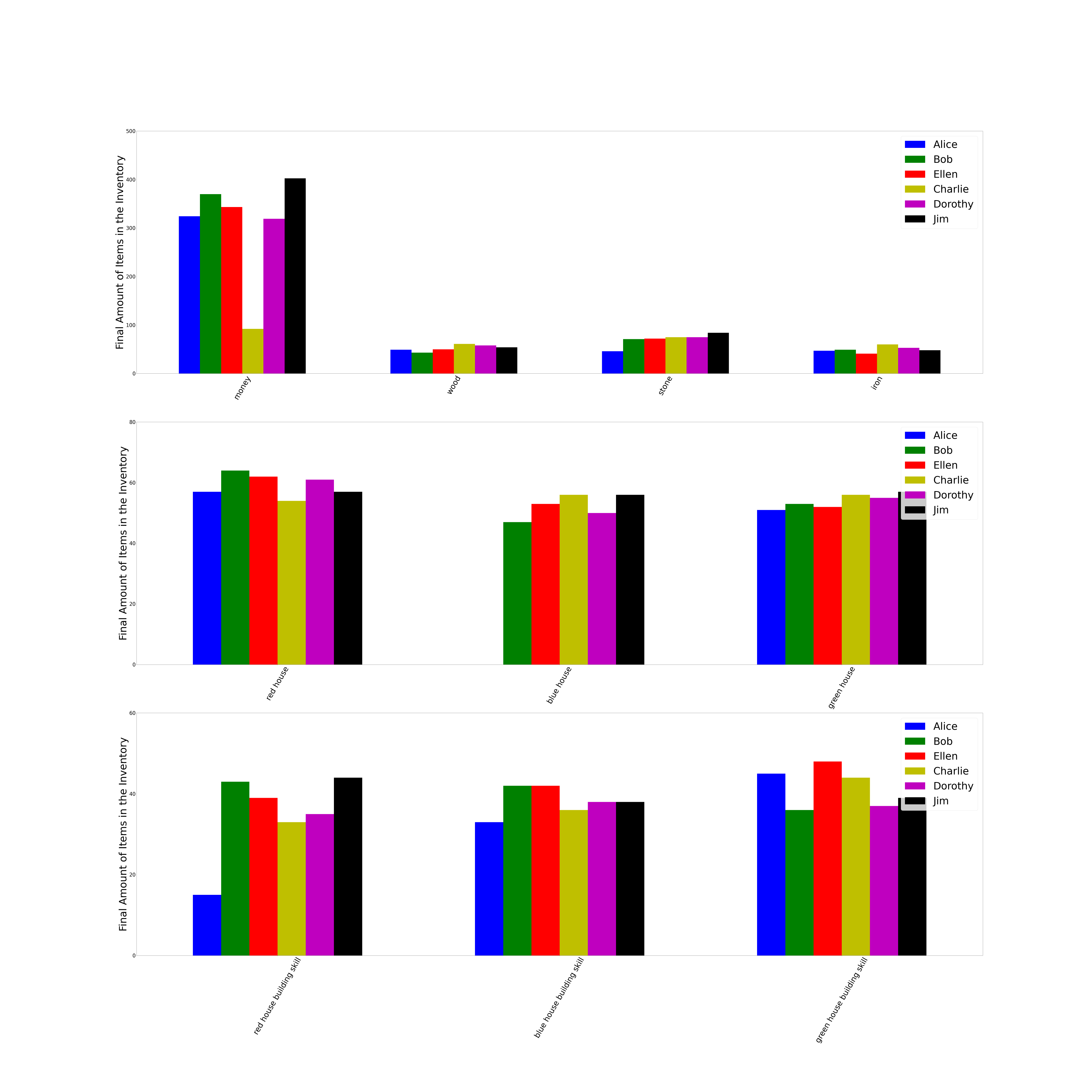}
	\caption{Final amounts of items in the inventories of six agents of the extended Concordia under Full-Utilitarian governing system when the game master cares about productivity in the society.}
	\label{Figure31}
\end{figure}

\end{document}